\pdfoutput=1
\documentclass[aps,prfluids,print,longbibliography]{revtex4-2}

\usepackage{newtxtext}
\usepackage{newtxmath}
\usepackage{ifpdf}
\usepackage{epstopdf}
\usepackage{graphicx}%figures and subfigures
\usepackage{subcaption}
\usepackage{caption}
\graphicspath{{figures/}}
\usepackage[export]{adjustbox}
\usepackage{amsmath}
\usepackage{mathtools}
\usepackage{bm}
\usepackage{multirow}
\usepackage{bigints}
\usepackage{xcolor}
\usepackage{tabularx}
\usepackage{enumitem}
\usepackage[%
  colorlinks=true,
  urlcolor=blue,
  linkcolor=blue,
  citecolor=blue
]{hyperref}
\usepackage{mathrsfs}
\usepackage{dblfloatfix}
\usepackage{float}

\def\mathbi#1{\textbf{\em #1}}

\def\Xint#1{\mathchoice
   {\XXint\displaystyle\textstyle{#1}}%
   {\XXint\textstyle\scriptstyle{#1}}%
   {\XXint\scriptstyle\scriptscriptstyle{#1}}%
   {\XXint\scriptscriptstyle\scriptscriptstyle{#1}}%
   \!\int}
\def\XXint#1#2#3{{\setbox0=\hbox{$#1{#2#3}{\int}$}
     \vcenter{\hbox{$#2#3$}}\kern-.5\wd0}}

\def\dashint{\Xint-}

\newcommand{\beq}{\begin{equation}}
\newcommand{\eeq}{\end{equation}}

\providecommand\bnabla{\boldsymbol{\nabla}}
\providecommand\bcdot{\boldsymbol{\cdot}}

\begin{document}

%\preprint{}

\title{Electrohydrodynamic instabilities in freely suspended viscous films under normal electric fields}


\author{Mohammadhossein Firouznia}
%\thanks{}
\author{David Saintillan}
\email[]{dstn@ucsd.edu}

\affiliation{Department of Mechanical and Aerospace Engineering, University of California San Diego, 9500 Gilman Drive, La Jolla, CA 92093, USA}


\date{\today}

\begin{abstract}
Electrohydrodynamic instabilities of fluid-fluid interfaces can be exploited in various microfluidic applications in order to enhance mixing, replicate well-controlled patterns or generate drops of a particular size. In this work, we study the stability and dynamics of a system of three superimposed layers of two immiscible fluids subject to a normal electric field. Following the Taylor-Melcher leaky dielectric model, the bulk remains electroneutral while a net charge accumulates on the interfaces. The interfacial charge dynamics is captured by a conservation equation accounting for Ohmic conduction, advection by the flow and finite charge relaxation. Using this model, we perform a linear stability analysis and identify different modes of instability, and we characterize the behavior of the system as a function of the relevant dimensionless groups in each mode. Further, we perform numerical simulations using the boundary element method in order to study the effect of nonlinearities on long-time interfacial dynamics. We demonstrate how the coupling of flow and surface charge transport in different modes of instability can give rise to nonlinear phenomena such as tip streaming or pinching of the film into droplets. 
\end{abstract}
%\keywords{}

\maketitle

\vspace*{-0.8cm}
\section{Introduction \label{sec:intro}}

The interface between two immiscible fluids can become unstable under the effect of an imposed electric field. In weakly conducting dielectrics, ion dissociation in the presence of an electric field is negligible, and therefore, diffuse Debye layers are absent in these systems where the fluid motion occurs as a result of the coupling between the electric and the hydrodynamic stresses at the interface. Following the Taylor-Melcher leaky dielectric model, the bulk of the fluid is assumed electroneutral and all free charges are concentrated on the interfaces separating the fluid volumes with different electrical properties \citep{Melcher_taylor1969LDM_annrev,saville1997annrevfluid}. The electric field acting on the interfacial charge creates electric stresses along the normal and tangential directions, which cause deformation and drag the fluid into motion. Surface tension acts mainly as a stabilizing effect, trying to restore the equilibrium interfacial shape.

In their pioneering works, Taylor and McEwan \citep{taylor1965stability_horiz_interafce} and Melcher and coworkers \citep{melcher_schwarz1968overstability,melcher1969EHD_horiz_interface} studied the effect of an electric field on the stability of fluid-fluid interfaces and provided analytical solutions in different limits.  Their analyses were followed by an extensive body of research aiming to obtain more accurate analytical solutions \citep{papageorgiou2004two_layer,li_Petropoulos2007two_layer_JFM,uguz2008POF_channel_two_layer}, to provide numerical solutions for large deformations \citep{thaokar2005POF_twolayer_channel,collins_Basaran2008nature_tipstreaming} and to study more complex configurations such as multi-layer systems \citep{Michael_Oneil1970three_layer_EHD,pease2002thin_film_linearstab,shankar2004film_EHD,Zahn_Reddy2006exp_theo_channel,zhang_lin2011JFM}. Of interest to us here is to understand the stability and dynamics of a system of superimposed fluid layers under the effect of an external electric field. Apart from its fundamental importance, there has been a renewed interest in this topic in recent years due to engineering applications where these electrohydrodynamic (EHD) instabilities have been exploited to develop techniques by which well-controlled patterns can be replicated on free surfaces, such as polymeric films or fluid layers flowing in microfluidic channels. In one of these techniques, EHD instabilities are employed to create fine periodic patterns in polymeric layers confined between a substrate and a mask \citep{chou1999lithographically,schaeffer_Ullrich2000Nature_LISA, schaffer_Ullrich2001EPL_EHD_polymerfilms,lin_Ulrich2002Macromolc_structure, Steiner2003Nature_hierarchical}. A summary of these applications is reviewed in \citep{Wu2009EHD_applications_review}. Theoretical analyses and numerical simulations have been employed to study the behavior of these systems under the following assumptions: (i) in the limit of zero inertia \citep{Thaokar_Kumaran2005POF_EHD_two_fluids,Craster_Matar2005POF_EHD_pattern_thinfilm, uguz2008POF_channel_two_layer}, (ii) in the limit of instantaneous charge relaxation where the charge is mainly transported via Ohmic current \citep{Thaokar_Kumaran2005POF_EHD_two_fluids,zhang_lin2011JFM}, and (iii) using the lubrication approximation for thin films \citep{pease2002thin_film_linearstab,shankar2004film_EHD,Thaokar_Kumaran2005POF_EHD_two_fluids,li_Petropoulos2007two_layer_JFM}.

More recently, EHD instabilities have been used in microfluidic devices to enhance mixing, or to generate droplets of a certain size. Zahn and Reddy \cite{Zahn_Reddy2006exp_theo_channel} developed a microfluidic mixer where two streams of immiscible fluids are mixed in a microchannel by applying an external field. Zhang \textit{et al.}~\cite{zhang_lin2011JFM} used transfer relations to study two- and trilayer systems in a variety of configurations such as unbounded geometries and channel flows, and provided analytical solutions in different limits of the Reynolds number. They recovered the previous analytical results of \cite{michael_Oneill_1970_EHD_3layer,papageorgiou2004generation} for trilayer systems, and those of \citep{uguz2008POF_channel_two_layer} for two-layers systems as special cases of their model. Considering the limit of instantaneous charge relaxation, they observed two modes of instability which they referred to as ``kink'' and ``sausage''.

In this work, we study the stability and dynamics of a freely suspended viscous film that is subject to a perpendicular external electric field. We use a charge transport model that incorporates finite charge relaxation, interfacial charge convection as well as Ohmic conduction from the bulk. This enables us to characterize the effect of charge convection on the behavior of the system, which is especially important under strong electric fields. We  present the governing equations in Sec.~\ref{sec:pb_def} and their non-dimensionalization in Sec.~\ref{sec:Nondim}. Next, we conduct a linear stability analysis in Sec.~\ref{sec:lin_stab} where we study the effect of the non-dimensional parameters governing the system on the fastest growing mode at the onset of instability. To supplement our theory, we employ boundary element simulations in Sec.~\ref{sec:Num_sim} to explore how the development of the flow and charge dynamics far from equilibrium gives rise to nonlinear phenomena such as tip streaming or pinching into droplets. Finally, we conclude and discuss the potential extensions to the present work in Sec.~\ref{sec:conclusion}. 

\section{Problem definition and governing equations\label{sec:pb_def}}
\begin{figure}[t]	
	\centering
	\adjustbox{trim={0.0\width} {0.0\height} {0.0\width} {0.0\height},clip}
	{\includegraphics[width=0.45
		\textwidth, angle=0]{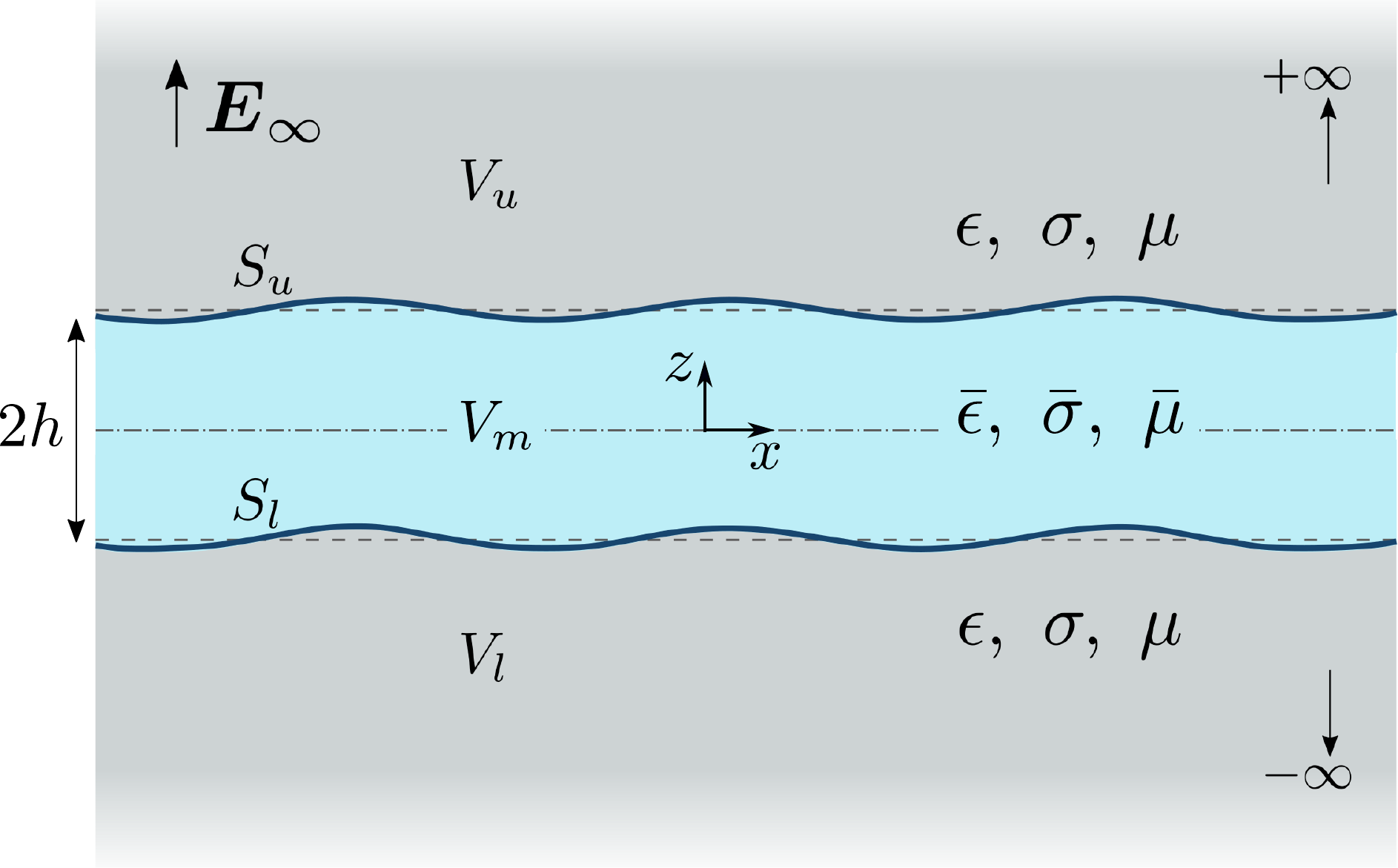} }	
	\caption{Problem definition: a liquid film suspended between two infinite immiscible liquid layers is subject to a perpendicular electric field $\bm{E}_{\infty}$. }  
	\label{figs:pb_def:schematics_3layer}  
\end{figure}

We study electrohydrodynamic instabilities that arise at the interfaces of a neutrally buoyant liquid film that is suspended in another liquid and subject to a uniform electric field along the perpendicular direction $\bm{E}_{\infty}=E_{\infty} \hat{\bm{e}}_z$. The liquid film occupying volume $V_m$ is surrounded by the upper layer $V_u$ from above, and by the lower layer $V_l$ from below as depicted in Fig.~\ref{figs:pb_def:schematics_3layer}. The subscripts $l$, $m$ and $u$ correspond to the lower, middle and upper layers, respectively. The interfaces separating the three liquid layers are denoted by $S_l$ and $S_u$. At equilibrium, the system is at rest, both interfaces are flat and coincide with the planes $z=\pm h$. We consider two-dimensional dynamics in the $(x,z)$ plane. The shape of each interface is parametrized as $z=\xi_{l,u}(x,t)$, with unit normal $\boldsymbol{n}$ pointing from the film into the suspending liquid.

The two phases are immiscible leaky dielectric Newtonian fluids with constant material properties. The electric permittivities, electric conductivities and viscosities are denoted by $(\bar{\epsilon},\bar{\sigma},\bar{\mu})$ in the film and by $(\epsilon,\sigma,\mu)$ in the suspending liquid. Following the Taylor-Melcher leaky dielectric model \cite{Melcher_taylor1969LDM_annrev}, the bulk is assumed to be electroneutral while any net charge in the system is concentrated on the interfaces between the two liquids. Consequently, the electric potential is harmonic in all layers: 
\begin{equation} \label{eq:intro:pot_laplace}
\nabla^2 \varphi_j(\bm{x})=0, \qquad   \bm{x}\in V_j, \qquad  j\in \{l,m,u\}. 
\end{equation} 
Far from the film, the electric field $\bm{E}=- \bnabla \varphi$ approaches the applied uniform field:
\begin{equation}
	\bm{E}\rightarrow \bm{E}_{\infty}=E_{\infty} \hat{\bm{e}}_z, \qquad \text{as} ~~ z\rightarrow \pm \infty. 
\end{equation} 
Across the interface, the tangential component of the electric field remains continuous while its normal component undergoes a jump due to the mismatch in electrical properties on both sides: 
\begin{equation}\label{eq:pb_def:ncrossE_Gausslaw}
\bm{n}\times\llbracket \bm{E} \rrbracket=0, \qquad  \bm{x}\in S_{j},~~j\in \{l,u\}.
\end{equation}
 As a result, a surface charge density $q(\bm{x})$ develops on each interface and is given by Gauss's law:
\begin{equation}
    q_i(\bm{x})=\bm{n} \bcdot \llbracket \epsilon \bm{E} \rrbracket, \qquad  \bm{x}\in S_{i},~~i\in \{l,u\}.
\end{equation}
 The jump operator introduced above is defined as the subtraction of any field variable on both sides of each interface, $\llbracket \mathcal{F} (\bm{x})  \rrbracket =\mathcal{F}_{l,u}(\bm{x}) - \mathcal{F}_{m}(\bm{x})$. The evolution of the surface charge density is described by a charge conservation equation accounting for finite charge relaxation, Ohmic conduction from the bulk and charge convection by the induced velocity: 
\begin{equation} 
	\partial_t q_i+\bm{n}\bcdot\llbracket \sigma \bm{E}\rrbracket + \bnabla_{\!s} \bcdot(q \bm{u})_i=0, \qquad \bm{x}\in S_{i},~~i\in \{l,u\}, \label{eq:pb_def:chgcons}
\end{equation}
where $\bnabla_{\!s}= (\bm{I}-\bm{n}\bm{n})\bcdot\bnabla$ is the surface gradient operator. 

Neglecting the effects of inertia and gravity, the fluid motion is governed by the Stokes equations in all layers:
	\begin{align}
	\bar{\mu}\nabla^2\bm{u}_m-\bnabla p_m&=\mathbf{0},    &  \bnabla \bcdot\bm{u}_m&=0,   & ~\bm{x}\in V_m,&\\
	\mu\nabla^2\bm{u}_{j}-\bnabla p_{j}&=\mathbf{0},     &  \bnabla \bcdot \bm{u}_{j}&=0,  & ~\bm{x}\in V_j, ~j\in \{l,u\}.&
    \end{align}
The velocity is continuous across the interfaces, and vanishes far away from the film:
\begin{eqnarray}
&	\llbracket \bm{u}(\bm{x}) \rrbracket=\bm{0}, \qquad  \bm{x}\in S_{l,u},& \label{eq:pb_def:cont_vel} \\
&\bm{u}(\bm{x}) \rightarrow \bm{0}, \qquad \text{as}~~  z \rightarrow \pm \infty.&\label{eq:pb_def:vel_infty}
\end{eqnarray} 
In the absence of Marangoni effects, the jump in electric and hydrodynamic tractions across each interface is balanced by surface tension forces:
\begin{equation}
	\llbracket \bm{f}^H\rrbracket + \llbracket \bm{f}^E\rrbracket=\gamma (\bnabla_{\!s}\bcdot \bm{n})\bm{n}, \qquad \bm{x}\in S_{l,u}, \label{eq:pb_def:dyn_BC}
\end{equation}
where $\gamma$ denotes the surface tension between the two liquids, assumed to be constant. Hydrodynamic and electric tractions can be expressed in terms of the Newtonian and Maxwell stress tensors, respectively: 
\begin{align}
 \bm{f}^H=\bm{n}\bcdot\bm{T}^H,&\qquad  \bm{T}^H=-p\bm{I}+\mu\big(\bnabla\bm{u}+{\bnabla\bm{u}}^T\big), \\ \label{eq:pb_def:jump_fE}
 \end{align}
 \begin{align}
 \bm{f}^E=\bm{n}\bcdot\bm{T}^E,& \qquad \bm{T}^E=\epsilon \left(\bm{E}\bm{E} -\tfrac{1}{2}E^2\bm{I}\right). \label{eq:pb_def:jump_fH}
 \end{align}
The electric traction can alternatively be expressed in terms of its tangential and normal components:
\begin{equation}
	\bm{f}^E =  \llbracket \epsilon E^n \rrbracket \bm{E}_t+\frac{1}{2} \llbracket \epsilon ({E^n}^2-{E^t}^2) \rrbracket \bm{n}=q \bm{E}_t+\llbracket p^E \rrbracket \bm{n}. \label{eq:pb_def:jump_fE_comps}
\end{equation}
The effect of the tangential electric field on the interfacial charge distribution is captured by the first term on the right-hand side. The second term captures normal electric stresses and can be interpreted as the jump in an electric pressure \cite{lac2007axisymmetric}.

Finally, the interfaces evolve and deform under the local velocity field as material surfaces. Defining the functions $g_{l,u}(\boldsymbol{x},t)=z-\xi_{l,u}(x,t)$, the kinematic boundary conditions read:
 \beq
 \dfrac{\mathrm{D}g_j}{\mathrm{D}t}=0, \qquad \bm{x}\in S_j, ~j\in \{l,u\},\label{eq:kinBC}
 \eeq
leading to the conditions
\begin{equation}
    \partial_t \xi_j = -u \partial_x \xi_j + w,\qquad \bm{x}\in S_j, ~j\in \{l,u\},
\end{equation}
where $\bm{u}=(u,w)$ are the velocity components. 
Also, the outward unit normal vector to the interface can be written as $\boldsymbol{n}_{l,u}=(\bnabla g/|\bnabla g|)_{l,u}$.

\section{Non-dimensionalization \label{sec:Nondim}}
For the system of governing equations presented above, dimensional analysis yields five non-dimensional groups, three of which characterize the mismatch in physical properties between the film and the suspending liquid: 
\beq
R=\frac{\sigma}{\bar{\sigma}}, \qquad Q=\frac{\bar{\epsilon}}{\epsilon},\qquad \lambda=\frac{\bar{\mu}}{\mu}.
\eeq 
A system with $\lambda>1$ corresponds to a film that is more viscous than the suspending liquid, and vice versa. The limit $\lambda\rightarrow \infty$ describes a rigid film, while $\lambda\rightarrow 0$ is relevant to describe a gas film suspended in a liquid. 

The other two non-dimensional groups can be obtained as the ratios of different times scales in the problem. First, we note that the conduction response of each liquid layer is characterized by the charge relaxation time scale, which is the time required for the free charge in the bulk to relax:
\begin{equation}
\tau_c=\frac{\epsilon}{\sigma}, \qquad   \bar{\tau}_c=\frac{\bar{\epsilon}}{\bar{\sigma}}. ~~ \label{eq:nondim:RQ}
\end{equation}
The product $RQ={\bar{\tau}_c}/{\tau_c}$ is the ratio of the charge relaxation time scales in both liquids. For instance, in the case $RQ<1$ conduction occurs at a faster rate in the film. Under an applied electric field, free charges in the bulk start to move towards the interfaces, resulting in the polarization of the film. This occurs on a time scale comparable to the Maxwell-Wagner relaxation time:
\begin{equation}
	\tau_{MW}=\frac{\bar{\epsilon}+2\epsilon}{\bar{\sigma}+2\sigma}= \dfrac{ \tau_c}{\alpha },\quad \text{where} \quad  \alpha=\frac{1+2R}{R(Q+2)}. \label{eq:nondim:tau_MW}
\end{equation} 

Following the accumulation of charge on the interfaces, the fluid is dragged into motion due to the force exerted by the electric field. This electrohydrodynamic flow deforms the interfaces on a time scale that is proportional to the inverse shear rate of the driving electric stress:
\begin{equation}
\tau_{EHD}=\frac{\mu(1+\lambda)}{\epsilon {E_{\infty}}^2}. \label{eq:nondim:tau_EHD}	
\end{equation}
In response, the interfacial tension acts as a restoring effect trying to minimize the surface area. The deformed interface recovers its equilibrium flat shape on the capillary time scale defined as:
\begin{equation}
	\tau_{\gamma}=\frac{\mu(1+\lambda)h}{\gamma}. \label{eq:nondim:tau_gamma}	
\end{equation}    
By comparing $\tau_c$, $\tau_{EHD}$ and $\tau_{\gamma}$, we can construct two additional non-dimensional groups. First, we define the electric capillary number as the ratio of the capillary time scale over the electrohydrodynamic flow time scale:
\begin{equation}
	Ca_E=\frac{\tau_{\gamma}}{\tau_{EHD}}=\frac{\epsilon {E_{\infty}}^2 h}{\gamma}. \label{eq:nondim:Ca}
\end{equation}      
According to Eq.~\eqref{eq:nondim:Ca}, the stronger the applied electric field the larger the electric capillary number. Moreover, the ratio of the charge relaxation time scale to the flow time scale defines the electric Reynolds number:
\begin{equation}
Re_E=\frac{\tau_c}{\tau_{EHD}}=\frac{\epsilon^2 {E_{\infty}}^2 }{\mu(1+\lambda)\sigma}, \label{eq:nondim:Re}
\end{equation} 
which characterizes the importance of charge convection versus conduction, two mechanisms responsible for the evolution of the interfacial charge distribution. 

We scale the governing equations and boundary conditions using time scale $\tau_{MW}$, length scale $h$, pressure scale $\epsilon E^2_{\infty}$, velocity scale $h \tau^{-1}_{EHD}$, and characteristic electric potential $h E_{\infty}$. The dimensionless momentum equations read: 
	\begin{align}\label{eq:nondim_stokes}
	&\nabla^2\bm{u}_m-(1+\lambda^{-1})\bnabla p_m=0,   \qquad  \bm{x}\in V_m,\\
	&\nabla^2\bm{u}_{j}-(1+\lambda)\bnabla p_{j}=0,  \qquad  \bm{x}\in V_{j},~~j\in \{l,u\}.
	\end{align}
The charge conservation equations become:
\begin{equation}\label{eq:nondim_charge_cons}
	\alpha \partial_t q_j+ \bm{n}\bcdot\big[   \bm{E}_j-R^{-1} \bm{E}_m \big] + Re_E \bnabla_s \bcdot(q\bm{u})_j=0,\qquad \bm{x}\in S_{j},~~j\in \{l,u\}
\end{equation}
where
\begin{equation}\label{eq:nondim_gausslaw}
q_j= \bm{n}\bcdot\left[\bm{E}_j- Q\bm{E}_m\right],\qquad \bm{x} \in S_{j},~~j\in \{l,u\}.
\end{equation}
%\begin{equation}
%\begin{split}
%&\bm{n}.\left[(\bm{E}_i\bm{E}_i -\frac{1}{2}E^2_i\bm{I} )-Q (\bm{E}_m\bm{E}_m -\frac{1}{2}E^2_m\bm{I}) \right] +  \\
%&\bm{n}.\left[(-p_i\bm{I}+\frac{ 1}{1+\lambda}(\nabla\bm{v}_i+{\nabla\bm{v}_i}^T)) -(-p_m\bm{I}+\frac{\lambda }{1+\lambda}(\nabla\bm{v}_m+{\nabla\bm{v}}_m^T)) \right]=\\
%& \frac{1}{Ca_E} (\nabla_s. \bm{n})\bm{n}, \qquad \text{for} ~~\bm{x}\in S_{i}, ~~i\in \{l,u\} 
%\end{split}	
%\end{equation}
The stress balance at the interfaces yields: 
\begin{align}
\begin{split}
\bm{n} \bcdot  \big[&-p_j\boldsymbol{I}+(1+\lambda)^{-1}(\bnabla\bm{u}_j+\bnabla\bm{u}_j^T) +p_m\boldsymbol{I}-(1+\lambda^{-1})^{-1}(\bnabla\bm{u}_m+{\bnabla\bm{u}}^T_m)\big] \\
& +\bm{n} \bcdot \big[(\bm{E}_j\bm{E}_j -\tfrac{1}{2}E_j^2\boldsymbol{I})-Q (\bm{E}_m\bm{E}_m -\tfrac{1}{2}E^2_m\boldsymbol{I})  \big]=Ca_E^{-1} (\bnabla_{\!s} \bcdot \bm{n})\bm{n},\quad \bm{x}\in S_{j}, ~~j\in \{l,u\}. \label{eq:dynamicBC}
\end{split}
\end{align}
Finally, the kinematic boundary conditions become:
\begin{equation}\label{eq:kinematicBC}
    \alpha \partial_t \xi_j = Re_E(w -u \partial_x \xi_j ),\qquad \bm{x}\in S_j, ~j\in \{l,u\}.
\end{equation}
In the remainder of the paper, all equations and variables are presented in non-dimensional form.

\section{Linear stability analysis \label{sec:lin_stab}}

\subsection{Theoretical formulation}

In this section, we perform a linear stability analysis to study the dynamic behavior of the system as a function of the governing parameters. In the base state (tilded variables), all liquid layers are at rest, both interfaces have flat shapes, $\tilde{\xi}_{u}=-\tilde{\xi}_{l}=1$, and the interfacial charges are $\tilde{q}_{u}=-\tilde{q}_{l}=1-RQ$. The applied electric field induces pressure jumps $\tilde{p}_u-\tilde{p}_m=\tilde{p}_l-\tilde{p}_m=(1-QR^2)/2$ across the interfaces due to the mismatch in electrical properties between the film and the suspending liquid. We consider infinitesimal perturbations (primed variables) applied to the base state variables:
\begin{align}
     &\xi_l=-1+\varepsilon \xi'_l,\quad \xi_u=1+\varepsilon \xi'_u,\quad q_l=\tilde{q}_l+\varepsilon q'_l,\quad q_u=\tilde{q}_u+\varepsilon q'_u, \label{eq:lin_stab:interface_pert}\\
     &\varphi_j=\tilde{\varphi}_j+ \varepsilon \varphi'_j, \quad \bm{u}_j=\tilde{\bm{u}}_j+ \varepsilon \bm{u}'_j, \quad p_j=\tilde{p}_j+ \varepsilon p'_j, \quad j\in\{l,m,u\}. \label{eq:lin_stab:field_pert} 
\end{align}
Next, we substitute Eqs.~\eqref{eq:lin_stab:interface_pert} and \eqref{eq:lin_stab:field_pert} into the governing equations and boundary conditions and linearize with respect to $\varepsilon$. The governing equations for the electric potential, velocity and pressure are: 
\begin{align}\label{eq:nondim_stokeslin}
    \nabla^2 \varphi'_j(\bm{x})=0, &    \qquad \bm{x}\in V_j,  \quad j\in \{l,m,u\},\\
	\nabla^2\bm{u}'_m-(1+\lambda^{-1})\bnabla p'_m=\mathbf{0},   &  \qquad \bm{x}\in V_m,\\
	\nabla^2\bm{u}'_{j}-(1+\lambda)\bnabla p'_{j}=\mathbf{0},  &  \qquad \bm{x}\in V_{j},  \quad j\in \{l,u\}.
	\end{align}
with jump conditions $\llbracket u'\rrbracket=\llbracket w'\rrbracket=0$ and $\hat{\bm{e}}_z\times\llbracket \bm{E}'\rrbracket+\bm{n}'\times \llbracket\tilde{\bm{E}}\rrbracket=\mathbf{0}$. The stress balance at the linearized location of the upper interface $z=1$ yields
\begin{eqnarray}
    	&(1-QR^2) \partial_x \xi_u' =  (1+\lambda^{-1})^{-1}\left(\partial_z u_m'+\partial_x w_m'\right) -(1+\lambda)^{-1} \left(\partial_z u_u'+\partial_x w_u' \right) +\partial_x \varphi_u'-RQ \partial_x \varphi_m',& \label{eq:lin_stab:stress_up_x} \\
    	&Ca_E^{-1} \partial_{xx} \xi'_u= 2(1+\lambda^{-1})^{-1} \partial_z w_m'-2(1+\lambda)^{-1} \partial_z w_u' +p'_u- p'_m + \partial_z \varphi_u'-RQ \partial_z \varphi_m' ,  &
\end{eqnarray}
along the $x$ and $z$ directions, respectively. The kinematic boundary condition, charge conservation equation and Gauss's law read: 
\begin{eqnarray}
 &\alpha	\partial_t \xi'_u=Re_E w'_u,& \\
&	\alpha \partial_t q_u' = \left[\partial_z \varphi_u' -R^{-1}\partial_z \varphi_m' \right] + Re_E(1-RQ)\partial_z w'_u,& \label{eq:lin_stab:q_dot_up} \\
  &  q_u'= Q \partial_z \varphi_m'-\partial_z \varphi_u' . &
\end{eqnarray}
Similarly, the linearized boundary conditions on the lower interface $z=-1$ are:
\begin{eqnarray}
   &(1-QR^2)\partial_x \xi'_l = (1+\lambda^{-1})^{-1}\left(\partial_z u'_m+\partial_x w'_m\right) -(1+\lambda)^{-1} \left(\partial_z u'_l+\partial_x w'_l \right)+ \partial_x \varphi'_l -RQ \partial_x \varphi'_m,	&\\
   &Ca_E^{-1} \partial_{xx} \xi'_l = 2(1+\lambda)^{-1} \partial_z w'_l -2(1+\lambda^{-1})^{-1} \partial_z w'_m +p'_m-p'_l+ RQ \partial_z \varphi'_m -\partial_z \varphi'_l ,&\\
   & \alpha	\partial_t \xi'_l=Re_E w'_l,&\\
   &\alpha \partial_t q'_l= \left[R^{-1} \partial _z\varphi'_m-\partial_z \varphi'_l \right]- Re_E(1-RQ)\partial_z w'_l,& \label{eq:lin_stab:q_dot_low}\\
   &q_l'= \partial_z \varphi_l'- Q \partial_z \varphi_m'.&
\end{eqnarray}
Next, we seek normal-mode solutions of the form $\varphi'(x,z,t)=\hat{\varphi}(z)\exp{(st+ikx)}$, with similar expressions for all the variables. $k$ is the wavenumber of the perturbation and $s$ is the corresponding growth rate. Substituting the normal-mode solutions into the governing equations, we obtain a  coupled system of differential equations for the normal-mode amplitude functions such as $\hat{\varphi}(z)$ (see appendix \ref{appendix:lin_stab} for details of the equations). Applying the boundary conditions along with the decay properties as $z\rightarrow \pm \infty$ results in an algebraic system for the unknown coefficients. Finally, we obtain the dispersion relation by setting the determinant of the algebraic system to zero.

Employing a more comprehensive charge transport model compared to the previous studies \cite{zhang_lin2011JFM,li_Petropoulos2007two_layer_JFM} results in a larger number of parameters. Therefore, the eigenvalue problem is solved numerically and the effect of each parameter on the stability of the system is characterized. Using our methodology for the case of two semi-infinite fluid layers as well as three-layer channel flow, we recover the analytical results of \cite{zhang_lin2011JFM} in the limit of instantaneous charge relaxation, where the charge is transported only via Ohmic conduction.

\subsection{Results and discussion\label{sec:lin_stab:results}}  

\begin{figure}[t]
	\centering
	\adjustbox{trim={0.0\width} {0.0\height} {0.0\width} {0.0\height},clip}
	{\includegraphics[width=0.9
		\textwidth, angle=0]{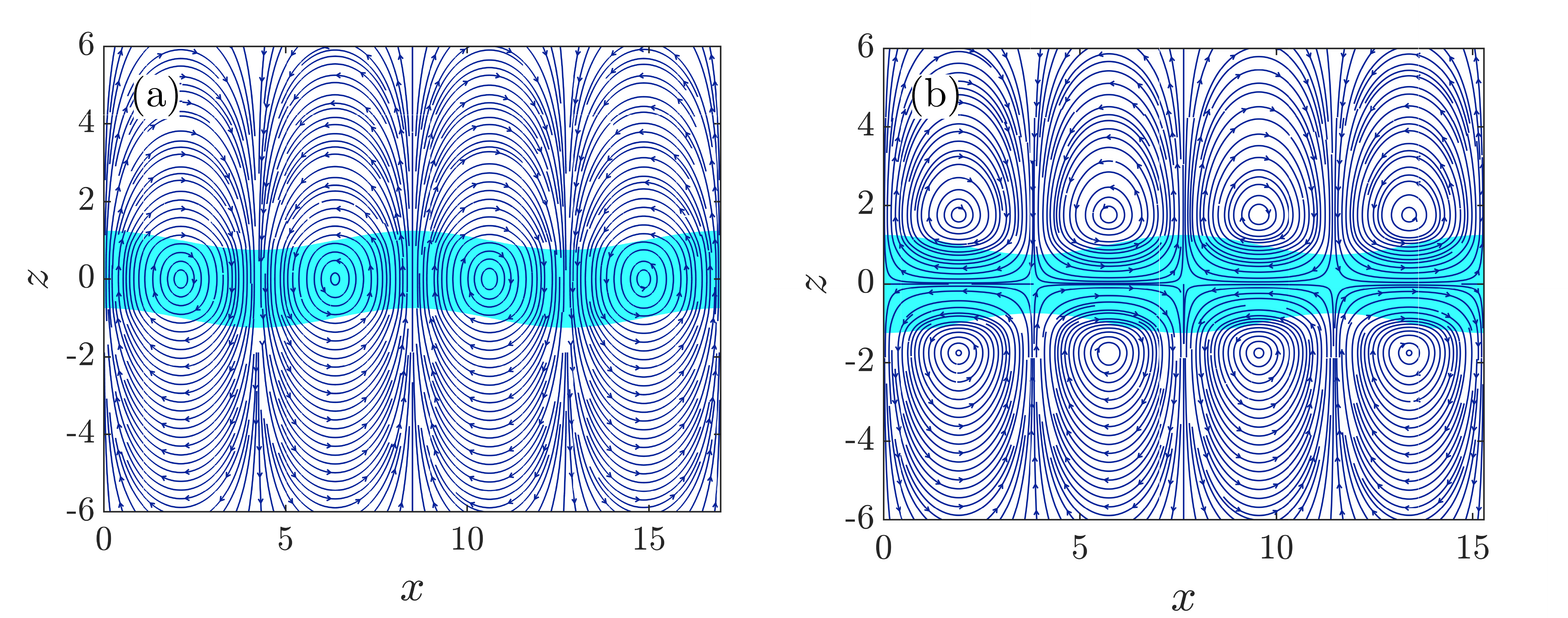} }
	\caption{Streamlines of the flow in the two dominant modes of linear instability: 
		(a) in-phase (sinuous) mode when $R=2$, (b) antiphase (varicose) mode when $R=0.5$. Other parameters are $(Q,\lambda,Ca_E,Re_E)=(1,1,10,1)$ in both systems. The blue region illustrates a typical shape of the film in the dominant mode.}  
	\label{figs:lin_stab:vec_streamlines} 
\end{figure}

The linear stability analysis (LSA) yields two dominant unstable eigenmodes. In the first mode, the instability is characterized by the growth of in-phase (sinuous) perturbations for the two interfaces ($\xi_u=\xi_l+2$). In the second mode, however, antiphase (varicose) perturbations are the most unstable ($\xi_u=-\xi_l$). Given that the dispersion equation is quadratic in $s$, there are two branches of solutions (eigenvalues) associated with each mode, one of which is dominant. Figure \ref{figs:lin_stab:vec_streamlines} shows the flow induced in each mode of instability along with the typical shapes of the film.

\begin{figure}[t]	
%	\centering
%	\adjustbox{trim={0.0\width} {0.0\height} {0.0\width} {0.0\height},clip}
	\includegraphics[width=0.9
		\textwidth, angle=0]{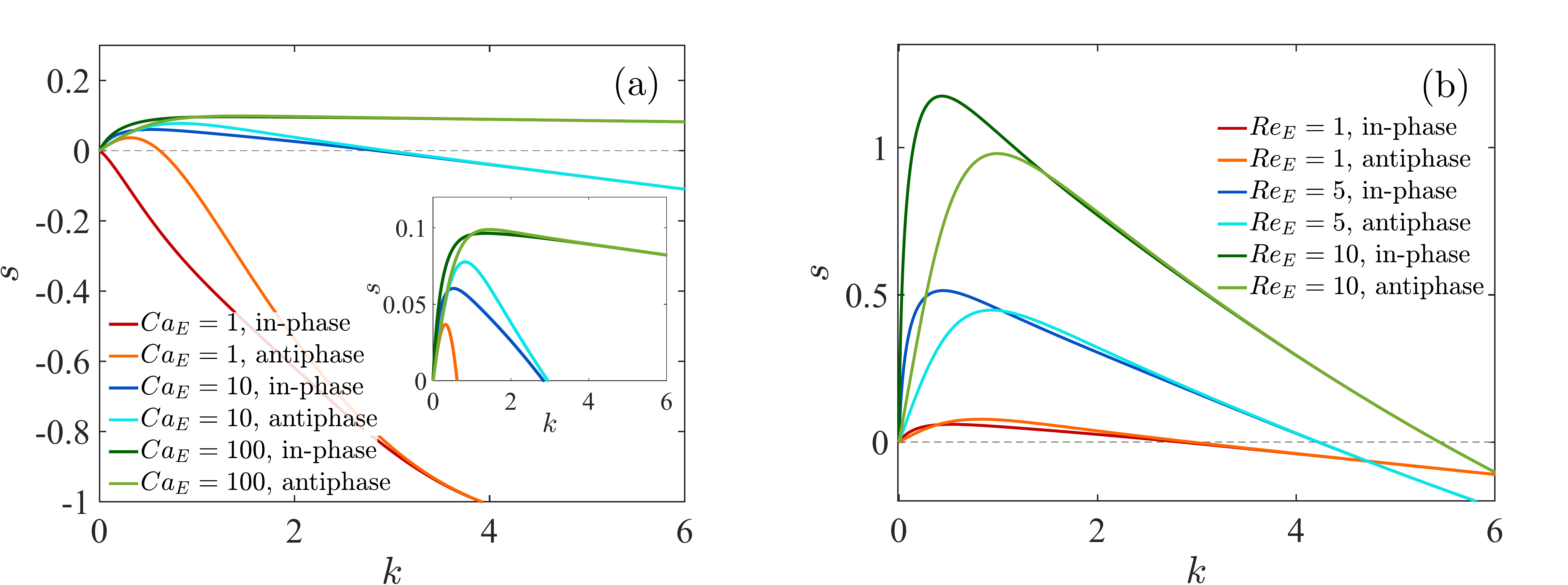} 
	\caption{Growth rate versus wavenumber in the dominant modes of linear instability for $(R,Q,\lambda)=(0.5,1,1)$: 
		(a) effect of electric capillary number when $Re_E=1$. Inset shows the unstable wavenumbers. (b) Effect of electric Reynolds number when $Ca_E=10$.}   
	\label{figs:lin_stab:s_vs_k_RQ05}
\end{figure}

Electric stresses acting on the interfacial charge can deform the film and drive the fluid into motion. The effect of the external electric field is characterized by both the electric capillary number and the electric Reynolds number. Figure \ref{figs:lin_stab:s_vs_k_RQ05} shows the growth rate as a function of wavenumber $k$ in each mode of instability for different values of $Ca_E$ and $Re_E$. For a given pair of leaky dielectric liquids, the electric capillary number $Ca_E$ characterizes the strength of the electric stresses versus surface tension effects. According to Fig.~\ref{figs:lin_stab:s_vs_k_RQ05}(a), the system is unstable at low wavenumbers over a finite range of $k$, and increasing $Ca_E$ destabilizes the system by increasing the maximum growth rate as well as the width of the unstable range. The electric Reynolds number plays a critical role in determining the charge transport regime and consequently the dynamics of the system. As $Re_E$ increases, the dominant mode of charge transport switches from Ohmic conduction to charge convection on the interface. According to Fig.~\ref{figs:lin_stab:s_vs_k_RQ05}(b), increasing $Re_E$ not only destabilizes the system, but can also alter the dominant mode of instability. 

\begin{figure}[t]	
	\centering
	\adjustbox{trim={0.0\width} {0.0\height} {0.0\width} {0.0\height},clip}
	{\includegraphics[width=0.8
		\textwidth, angle=0]{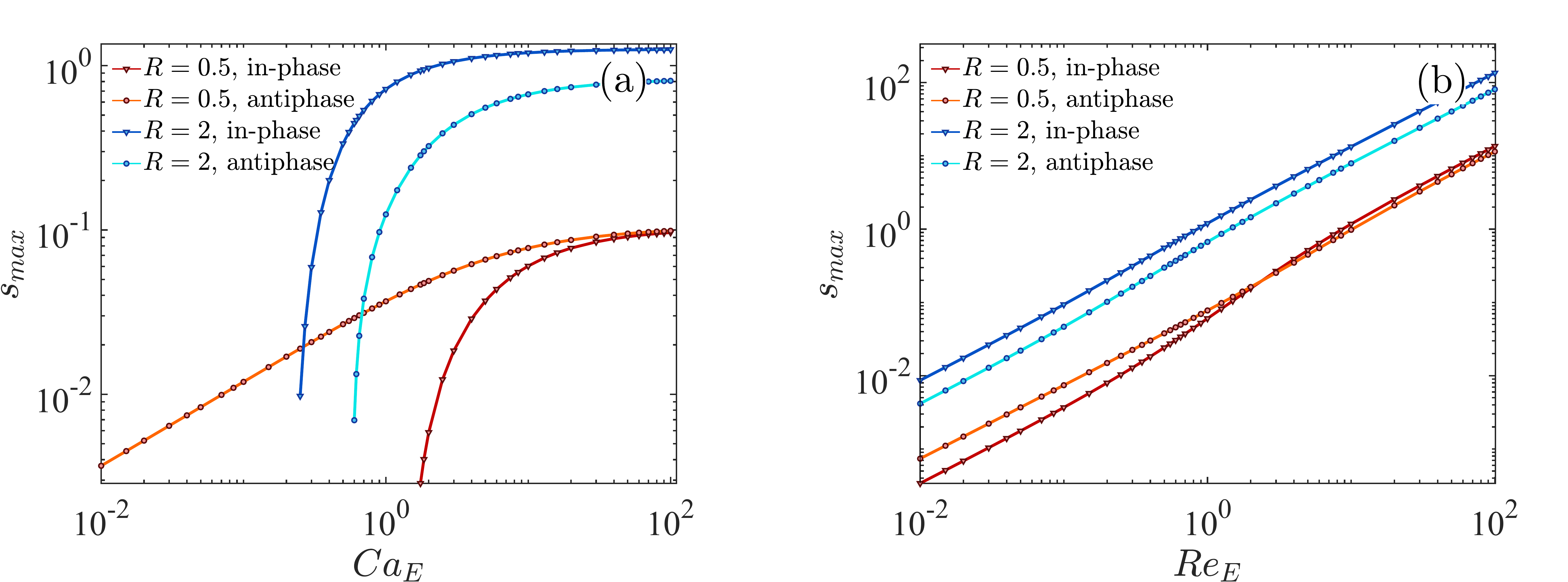} }	
	\caption{Maximum growth rate in each mode of instability as a function of: (a) electric capillary number, and (b) electric Reynolds number for two different systems with $R=0.5$ and $R=2$ while $(Q,\lambda)=(1,1)$. $Re_E=1$ in (a) and $Ca_E=10$ in (b).}  
	\label{figs:lin_stab:smax_vs_Ca_Re}  
\end{figure}

The fate of the system following the onset of instability is determined by the dynamical behavior in the fastest growing mode. The maximum growth rate and the corresponding wavenumber in each mode are denoted respectively as $s_{max}$ and $k_{max}$. We studied the effect of each non-dimensional group on the maximum growth rate in Figs.~\ref{figs:lin_stab:smax_vs_Ca_Re} and \ref{figs:lin_stab:smax_vs_lambda_R_Q} (for more information on $k_{max}$ see appendix \ref{appendix:kmax_plots}). Information about the fastest growing mode is of practical importance for engineering applications where electrohydrodynamic instabilities may be utilized to generate interfacial patterns with a prescribed length scale  \cite{chou1999lithographically,Steiner2003Nature_hierarchical, Zahn_Reddy2006exp_theo_channel}. 
When $RQ<1$ (${\bar{\tau}_c}<{\tau_c}$), the charge is provided to the interface at a faster rate in the film than in the bulk. Therefore, the interfacial charge is predominantly provided from the film, and the dipole moment is aligned with the external electric field. Conversely, when $RQ>1$, the suspending liquid is more conducting and the dipole moment is anti-parallel to the applied electric field. Since displacing the dipole moment by deforming the interface results in a destabilizing torque, the system is inherently more unstable in this configuration. This is confirmed in Figs.~\ref{figs:lin_stab:smax_vs_Ca_Re} and \ref{figs:lin_stab:smax_vs_lambda_R_Q}(a) when comparing the maximum growth rates between two model systems with $RQ=2$ and $0.5$.  We note the two systems are comprised of identical leaky dielectric liquids that are arranged in an opposite order in the film and the suspending phase ($R_1=R^{-1}_2$). According to Fig.~\ref{figs:lin_stab:smax_vs_Ca_Re}(a), the maximum growth rate in each mode increases with $Ca_E$ until it plateaus at $Ca_E\gg1$. Figure \ref{figs:lin_stab:smax_vs_Ca_Re}(b) shows the maximum growth rate as a function of $Re_E$ for two different configurations. Increasing the electric Reynolds number has a destabilizing effect in both cases. Additionally, the mode of instability (in-phase vs antiphase) can switch as a function of electric Reynolds number. The maximum growth rate is found to scale linearly with $Re_E$ in both limits of $Re_E\gg1$ and $Re_E\ll1$. 
%Furthermore, in-phase perturbations are most unstable at large $Re_E$, the behavior which is not limited to these two model systems. Based on equations \eqref{eq:lin_stab:q_dot_up} and \eqref{eq:lin_stab:q_dot_low}, at $Re_E\gg1$ the charges are transported on both interfaces predominantly via charge convection:
%\begin{align}
%    & \partial_t q_u' \approx  Re_E(1-RQ)\alpha^{-1} \partial_z w_u',\\
%    & \partial_t q_l' \approx  -Re_E(1-RQ)\alpha^{-1} \partial_z w_l',
%\end{align}
%where $\partial_z w_u'=-\partial_z w_l'$ based on the decay properties as $z$. Therefore, $\partial_t q_u'=\partial_t q_l'$ which means the interfacial charges evolve in-phase on both interfaces.

The effect of the ratios of material properties is considered in more detail in Fig.~\ref{figs:lin_stab:smax_vs_lambda_R_Q}. We observe in Fig.~\ref{figs:lin_stab:smax_vs_lambda_R_Q}(a) that a large viscosity ratio (more viscous film) is unfavorable for stability regardless of the configuration. Moreover, the dominant mode of stability switches from antiphase to in-phase when going from small to large values of $\lambda$. Figures \ref{figs:lin_stab:smax_vs_lambda_R_Q}(b) and (c) characterize the effect of conductivity ratio $R$ and permittivity ratio $Q$ on the stability of the system, while  all other material properties are kept the same between the fluid layers. The larger growth rates observed at large values of $R$ and $Q$ confirm our expectation that $RQ>1$ (${\bar{\tau}_c}>{\tau_c}$) corresponds to a system that is more electrically unstable.  According to Fig.~\ref{figs:lin_stab:smax_vs_lambda_R_Q}(c), the  maximum growth rate is independent of the permittivity ratio when $Q\ll1$ and the mode of instability changes from antiphase to in-phase going from $Q<1$ to $Q>1$. We also note that the system becomes stable as $RQ$ approaches $1$ in both Fig.~\ref{figs:lin_stab:smax_vs_lambda_R_Q}(b) and (c) since $RQ=1$  corresponds to a non-polarizing system in the electric field.

\begin{figure}[t]	
	\centering
	\adjustbox{trim={0.0\width} {0.0\height} {0.0\width} {0.0\height},clip}
	{\includegraphics[width=1
		\textwidth, angle=0]{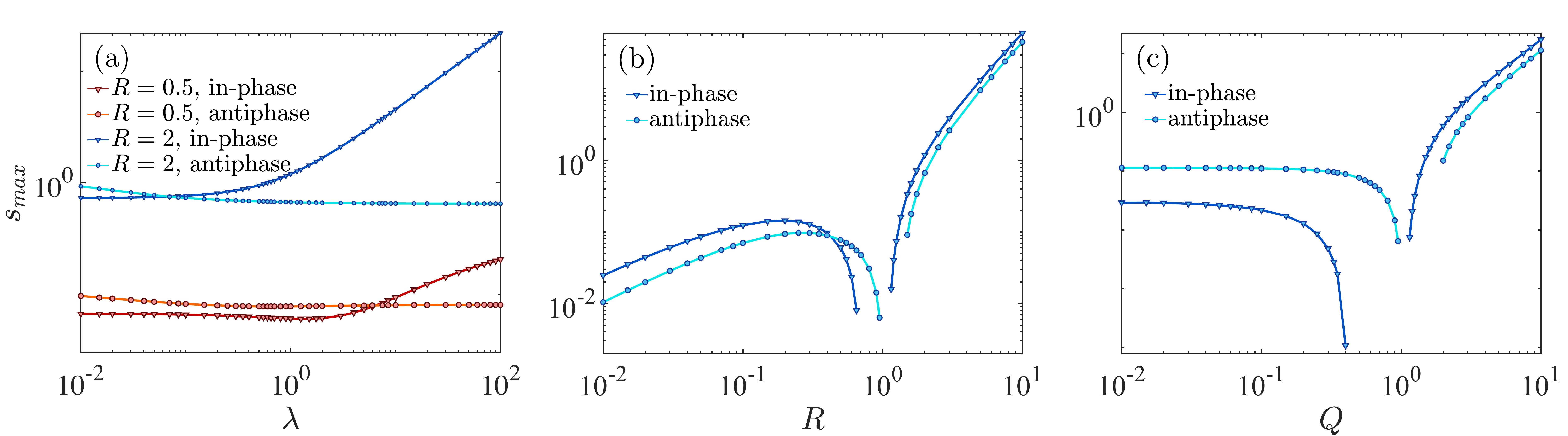} }%	smax_lambda_7March2021
	\caption{Maximum growth rate in each mode of instability as a function of: (a) viscosity ratio $\lambda$ for two different systems with $R=0.5$ and $R=2$, (b) conductivity ratio $R$, and (c) permittivity ratio $Q$. The permittivity ratio is set to $Q=1$ in (a,b), the conductivity ratio is $R=1$ in (c) and $\lambda=1$ in (b,c). In all systems $(Ca_E,Re_E)=(10,1)$.  }  
	\label{figs:lin_stab:smax_vs_lambda_R_Q}  
\end{figure}

%\begin{figure}[t]	
%	\centering
%	\adjustbox{trim={0.0\width} {0.0\height} {0.0\width} {0.0\height},clip}
%	{\includegraphics[width=0.9
%		\textwidth, angle=0]{./linear_stability/NEW_KBC/smax_vs_R_Q} }	
%	\caption{Maximum growth rate in each mode of instability as a function of: (a) conductivity ratio $R$, and (b) permittivity ratio $Q$ for a system with $(\lambda,Ca_E, Re_E)=(1,10,1)$. The permittivity ratio is set to $Q=1$ in (a), while the conductivity ratio is $R=1$ in (b).}  
%	\label{figs:lin_stab:smax_vs_R_Q}  
%\end{figure}

%In Sec.\ref{sec:Num_sim}, we turn to the dynamics of the system in the nonlinear regime of growth, which we study using boundary element simulations. We demonstrate how the coupling of the flow and the charge dynamics in each mode of instability gives rise to nonlinear phenomena such as tip-streaming and pinching into droplets. 

\section{Numerical simulations \label{sec:Num_sim}}
In this section, we complement the linear stability analysis of Sec.~\ref{sec:lin_stab} with numerical simulations. We present a numerical method in Sec.~\ref{sec:Num_sim:BEM_formul} for the nonlinear solution of the system of governing equations \eqref{eq:nondim_stokes}--\eqref{eq:kinematicBC} based on the boundary integral equation for Laplace and Stokes equations in a periodic domain of period $L_p$ along the $x$-direction. The size of the domain is chosen based on the wavenumber associated with the fastest growing mode obtained via linear stability analysis: $L_p=2\pi k_{max}^{-1}$. These simulations provide us with insight into the dynamic behavior of the system beyond the linear regime of instability. Our methodology shares similarities with that of \citep{Das2017JFM_small_deformation, Das2017EHD_drop_numerical,Hossein2021JFMdraft} for interfaces separating two liquid layers. We implement adaptive grid refinement to handle large local deformations, curvature and charge gradients in the nonlinear regime of growth. Results are presented in Sec.~\ref{sec:Num_sim:results} where we compare the predictions from the linear theory (LSA) of Sec.~\ref{sec:lin_stab} and from numerical simulations (NS). Nonlinear dynamics are also explored in transient simulations far past the onset of instabilities, where the interplay between charge dynamics and fluid flow gives rise to nonlinear phenomena such as tip streaming and pinching into drops.

\subsection{Boundary element method \label{sec:Num_sim:BEM_formul}}
Laplace's equation \eqref{eq:intro:pot_laplace} for the electric potential can be reformulated as a single-layer integral equation \cite{sherwood1988JFM_breakup, Hstone1998EHD_droppairs, lac2007axisymmetric}:
\begin{equation}\label{eq:BEM:potential_SLP}
	\varphi_{l,m,u}( \bm{x}_0 )=-\bm{x}_0\bcdot\bm{E}_{\infty} -\int_{S} \bm{n}\bcdot\llbracket \nabla \varphi( \bm{x} ) \rrbracket \mathcal{G}^P \left( \bm{x}_0;\bm{x} \right) \mathrm{d}l(\bm{x}),\qquad \text{for}~ \bm{x}_0\in V, S	,
\end{equation}
where $V=V_l\cup V_m \cup V_u,$ and $S=S_l\cup S_u$. The evaluation point $\bm{x}_0$ can be anywhere in space while the integration point $\bm{x}$
  lies on one of the two interfaces. The periodic Green's function for Laplace's equation $\mathcal{G}^P$ represents the potential due to a periodic array of point sources separated by distance $L_p$ along the $x$ axis \cite{pozrikidis2011introduction, pozrikidis_book2002BEMpractical}: 
 \begin{equation}\label{eq:BEM:Green_laplace_2DP}
 	\mathcal{G}^P(\bm{x};\bm{x}_0)=-\frac{1}{4\pi}\ln{\bigg[2 \Big\{ \cosh{\left[k_p(y-y_0)\right]}-\cos{\left[ k_p(x-x_0)\right]} \Big\} \bigg]}  ,
 \end{equation}
 where $k_p=2\pi/L_p$ is the wavenumber associated with the geometrical periodicity. Taking the gradient of Eq.~\eqref{eq:BEM:potential_SLP} with respect to $\bm{x}_0$, we obtain an integral equation for the electric field: 
 \begin{equation}
 	\bm{E}_{l,m,u}( \bm{x}_0 )=\bm{E}_{\infty} -\int_{S} \llbracket E^n( \bm{x} ) \rrbracket \bnabla_0 \mathcal{G}^P dl(\bm{x}),\qquad \text{for}~ \bm{x}_0\in V. \label{eq:BEM:SLP_E}
 \end{equation}
The derivative of the Green's function undergoes a discontinuity across each interface \cite{pozrikidis2011introduction}. Therefore, we express the electric field on either side of each interface as: 
  \begin{align}
  	\bm{E}_{l,u}( \bm{x}_0 )=\bm{E}_{\infty} -\dashint_{S}\llbracket E^n( \bm{x} ) \rrbracket \nabla_0 \mathcal{G}^P \mathrm{d}l(\bm{x})+\dfrac{1}{2}\llbracket E^n( \bm{x}_0 ) \rrbracket \bm{n}(\bm{x}_0),\qquad \text{for}~ \bm{x}_0\in S, \label{eq:BEM:SLP_grad_out}\\
  \bm{E}_{m}( \bm{x}_0 )=\bm{E}_{\infty} -\dashint_{S} \llbracket E^n( \bm{x} ) \rrbracket \nabla_0 \mathcal{G}^P \mathrm{d}l(\bm{x})-\dfrac{1}{2}\llbracket E^n( \bm{x}_0 ) \rrbracket \bm{n}(\bm{x}_0),\qquad \text{for}~ \bm{x}_0\in S. \label{eq:BEM:SLP_grad_in}
  \end{align}
  The second terms on the right-hand side of \eqref{eq:BEM:SLP_grad_out} and \eqref{eq:BEM:SLP_grad_in} denote the principal-value integral where the evaluation point is located precisely on the interfaces. The singularity in Eqs.~\eqref{eq:BEM:SLP_grad_out} and \eqref{eq:BEM:SLP_grad_in} can be removed by taking a dot product with the normal vector $\bm{n}(\bm{x}_0)$. Finally, we obtain the following equation after combining the results with Gauss's law Eq.~\eqref{eq:nondim_gausslaw}:   
  \begin{equation}
  	\dashint_{S} \llbracket E^n( \bm{x} ) \rrbracket [\bm{n}(\bm{x}_0)\bcdot\bnabla_0 \mathcal{G}^P] \mathrm{d}l(\bm{x})-\dfrac{1+Q}{2(1-Q)}\llbracket E^n( \bm{x}_0 ) \rrbracket =E^n_{\infty}( \bm{x}_0 )-\dfrac{q(\bm{x}_0)}{1-Q},\qquad \text{for}~ \bm{x}_0\in S, \label{eq:BEM:integral_eq_En}
  \end{equation}
which is an integral equation for $\llbracket E^n \rrbracket$ as a function of the charge distribution. Subsequently, we determine the normal component of the electric field at each interface based on Gauss's law:
\begin{equation}
	E^n_{l,u}=\dfrac{q-Q\llbracket E^n \rrbracket}{1-Q}, \qquad  E^n_{m}=\dfrac{q-\llbracket E^n \rrbracket}{1-Q}. \label{eq:BEM:integral_eq_Eout}
\end{equation}  
The electric potential is obtained via \eqref{eq:BEM:potential_SLP} in the next step. This allows us to obtain the tangential component of the electric field by numerically differentiating the electric potential along $S$. This way, we avoid using \eqref{eq:BEM:SLP_grad_out} and \eqref{eq:BEM:SLP_grad_in}, which are singular and require further treatment \cite{sellier2006singularity_BEM}. Having both components of the electric field at the interface, we can find the jump in electric tractions $\llbracket \bm{f}^E \rrbracket$ from \eqref{eq:pb_def:jump_fE_comps}. The jump in the hydrodynamic tractions can then be obtained using the interfacial stress balance \eqref{eq:pb_def:dyn_BC}. Finally, the interfacial velocity is determined using the Stokes boundary integral equation, in its dimensionless form \cite{rallison1978numerical,pozrikidis1992BEM_redbook}:
 \begin{equation}
 \bm{u}(\bm{x}_0) = -\dfrac{1}{2\pi} \int_{S}   \llbracket \bm{f}^H(\bm{x}) \rrbracket  \bcdot \bm{G}^P(\bm{x};\bm{x}_0)\,\mathrm{d}l(\bm{x}) 
 + \dfrac{1-\lambda}{2\pi (1+\lambda)}\,\, \dashint_{S}\bm{u}(\bm{x})\bcdot\bm{T}^P(\bm{x};\bm{x}_0) \bcdot \bm{n}(\bm{x})\mathrm{d}l(\bm{x}),  \quad \text{for}~ \bm{x}_0\in S,
 \label{eq:BEM:integral_eq_Stokes}
 \end{equation}
where $\bm{G}^P$ is the singly periodic Green's function describing the flow due to a periodic array of point forces separated by the distance  $L_p$ along the $x$ direction, and $\bm{T}^P$ is the corresponding stress tensor \citep{pozrikidis1992BEM_redbook}. Equation \eqref{eq:BEM:integral_eq_Stokes}, which is a Fredholm integral equation of the second kind for $\bm{u}$, yields a dense linear system after discretization. The linear algebraic system is solved iteratively using a GMRES method \cite{GMRES_saad1986SIAM, GMRES_code1997set}, and the interfacial velocity is used to update the charge distribution and the shape of each interface via the kinematic boundary conditions. The numerical algorithm during one time step of the simulations can be summarized as follows:   
\begin{enumerate}[label=(\Roman*)]
     \item Calculate $\llbracket E^n \rrbracket$, $E^n_{l,u}$ and $E^n_m$ for the given charge distribution $q(\bm{x})$, by solving the integral equation \eqref{eq:BEM:integral_eq_En} along with \eqref{eq:BEM:integral_eq_Eout}.
     \item Compute the electric potential on each interface using \eqref{eq:BEM:potential_SLP}.
     \item Differentiate the surface potential numerically along each interface in order to obtain the tangential electric field $\bm{E}^t=-\bnabla_{\!s} \varphi$. 
     \item Knowing both components of the electric field on each interface, evaluate the jump in electric tractions $\llbracket \bm{f}^E \rrbracket$ via \eqref{eq:pb_def:jump_fE_comps} and use it to determine the jump in hydrodynamic tractions $\llbracket \bm{f}^H \rrbracket$ via \eqref{eq:pb_def:dyn_BC}.
     \item Solve for the interfacial velocity by inverting the discretized Stokes boundary integral equation \eqref{eq:BEM:integral_eq_Stokes}. 
     \item Update the charge distribution by integrating \eqref{eq:nondim_charge_cons} explicitly in time using a second-order Runge-Kutta scheme.
     \item Advance the position of both interfaces by advecting the grid using the normal component of the interfacial velocity: $\dot{\bm{x}}_i(t)=(\bm{u}\bcdot\bm{n})\bm{n}$. 
     \item Refine the grid locally if either the local curvature, element length or magnitude of the charge gradient exceed certain thresholds.        
\end{enumerate}

We use piecewise cubic spline interpolation to represent the shape of the interface with continuous slope and curvature from one element to another. This provides an easy and accurate way to compute the geometrical properties such as curvature, normal and tangent vectors. In case either the mean curvature, the length or the magnitude of the charge gradient in an element exceeds the predefined thresholds, that element is divided into two new elements. The position of the new node along with other variables such as velocity and charge density are evaluated using spline interpolation. The use of adaptive grid refinement significantly improved the performance of our numerical scheme in handling large local deformations and charge gradients in the nonlinear regime. Nonetheless, the number of new elements is only a fraction of total elements in each iteration. Excessive grid refinement can result in spurious oscillations in the spline representation and excite numerical instabilities \cite{dBoor1978splines_book}.

\subsection{Results and discussion\label{sec:Num_sim:results}}

 \begin{table}[b] %add [H] placement to break table across pages
 	\centering
 	\caption{Different systems studied using numerical simulations and their dimensionless parameters. \label{table:list_systems}}
 	\begin{ruledtabular}
 		\begin{tabular}{c c c c c c}
 			system & $R$ & $Q$ & $\lambda$ & $Ca_E$ & $Re_E$ \\[0.5ex] 
 			\hline
 			$\mathbi{S1}$  & $0.5$ & $1$ & $1$ & $10$ & $1$ \\
 			$\mathbi{S2}$ & $2$ & $1$ & $1$ & $10$ & $1$ \\
 			$\mathbi{S3}$ & $-$ & $0.1$ & $1$ & $10$ & $1$ \\
 			$\mathbi{S4}$ & $1$ & $-$ & $1$ & $10$ & $1$ \\
 			$\mathbi{S5}$ & $0.5$ & $1$ & $-$ & $10$ & $1$ \\
 			$\mathbi{S6}$ & $-$ & $1$ & $1$ & $10$ & $1$ \\
 			$\mathbi{S7}$ & $2$ & $1$ & $-$ & $10$ & $1$ \\
 		\end{tabular}
 	\end{ruledtabular}
 \end{table}
 
  \begin{figure}[t]
	\centering
	\adjustbox{trim={0.0\width} {0.0\height} {0.0\width} {0.0\height},clip}
	{\includegraphics[width=0.75
		\textwidth, angle=0]{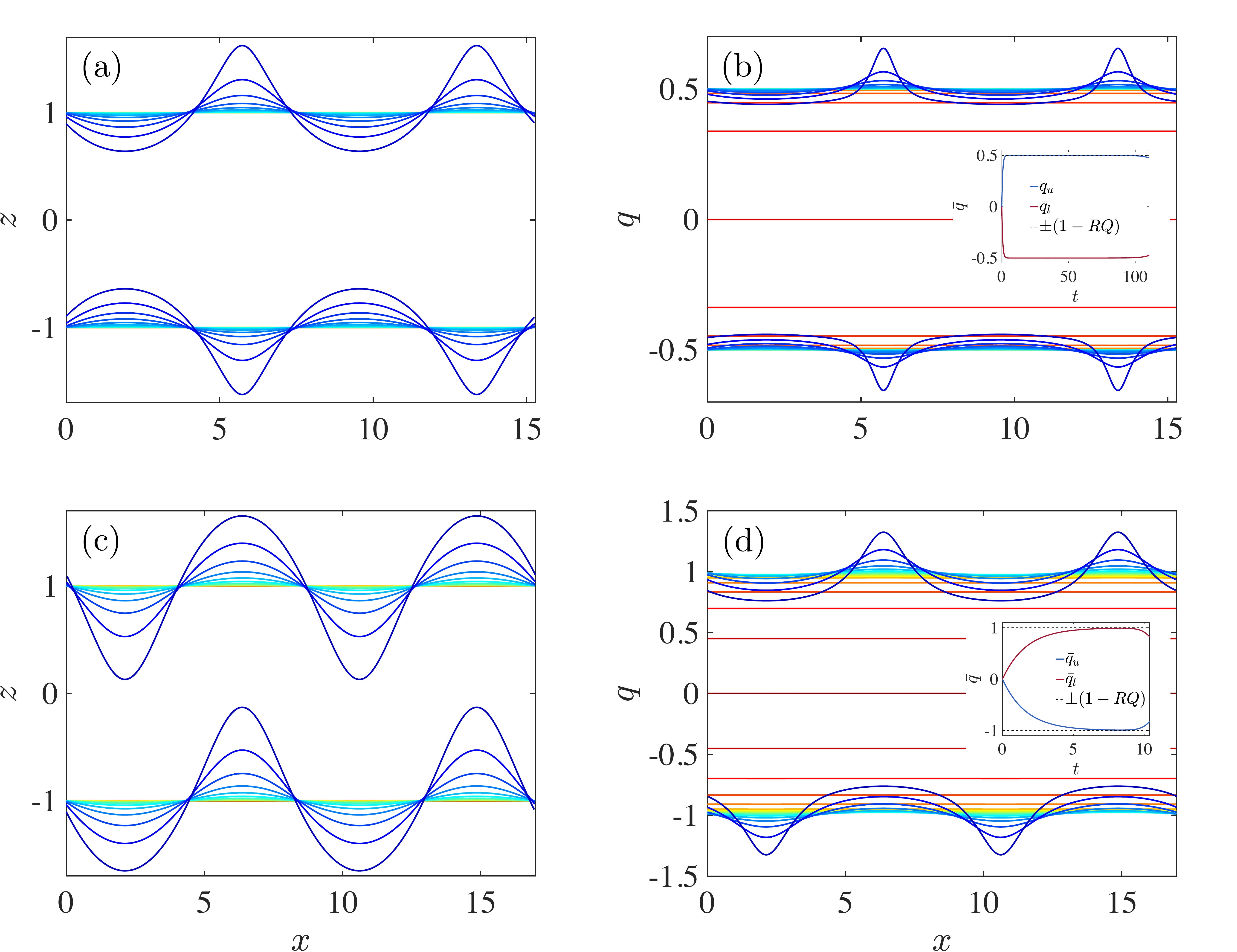} }	
	\caption{Evolution of the shape of both interfaces (left column) and their charge distributions (right column) in different modes of instability: 
		(a,b) antiphase mode for system $\mathbi{S1}$, (c,d) in-phase mode for system $\mathbi{S2}$.  Insets in (b,d) show the average charge densities $\bar{q}$ on each interface as a function of time. In the insets, red and blue curves show the transient average charge on the upper and lower interface, respectively, while the dashed lines show the steady interfacial charges in the base state. }  
	\label{figs:BEM:time_lapse_qavg}  
\end{figure}    

\begin{figure}[h]
	\centering
	\adjustbox{trim={0.0\width} {0.0\height} {0.0\width} {0.0\height},clip}
	{\includegraphics[width=0.75
		\textwidth, angle=0]{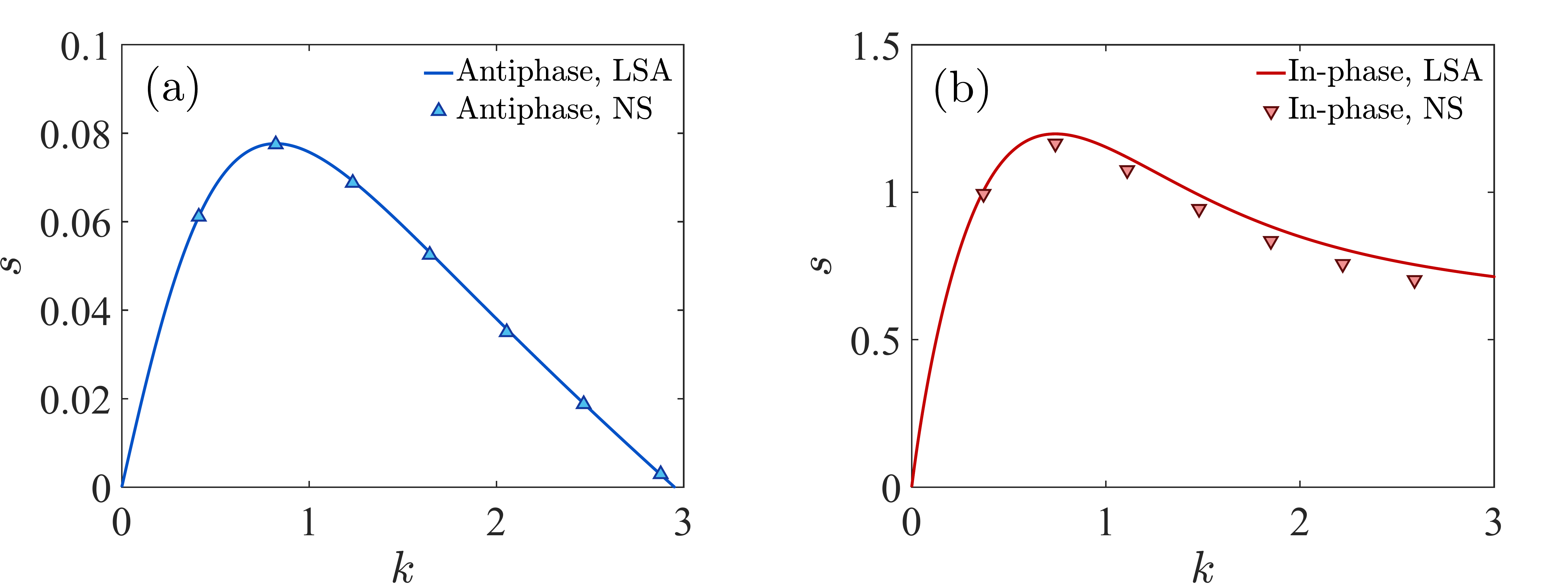} }	
	\caption{Growth rate $s$ as a function of wavenumber $k$ in the dominant mode of instability obtained via numerical simulations (NS) and linear stability analysis (LSA) for $\mathbi{S1}$ in (a), and $\mathbi{S2}$ in (b). In these simulations, the size of the domain is set to $L_p=4\pi k_{max}^{-1}$, which is twice the wavelength associated with the fastest growing mode. }  
	\label{figs:BEM:s_LSA_vs_BEM}  
\end{figure}

The list of systems considered in our simulations is provided in Table \ref{table:list_systems} along with the corresponding parameter values. We first consider in Fig.~\ref{figs:BEM:time_lapse_qavg} two representative cases, $\mathbi{S1}$ and $\mathbi{S2}$, that exhibit antiphase and in-phase instabilities, respectively. Initially, both interfaces are uncharged with a flat shape. The film polarizes mainly via Ohmic conduction at short times, and the charge distribution first increases uniformly on both interfaces as shown in Fig.~\ref{figs:BEM:time_lapse_qavg}(b,d). Meanwhile, perturbations of certain modes start to grow until the system becomes unstable, consistent with our predictions from LSA. The growth rates obtained in our simulations are in a close agreement with the theoretical values from LSA as shown in Fig.~\ref{figs:BEM:s_LSA_vs_BEM}. Following the onset of instability, as the induced flow becomes stronger, the effect of charge convection becomes more significant. The accumulation of charge on certain points on each interface results in large local electric stresses exerted by the applied electric field, which further deforms the interface as shown in Fig.~\ref{figs:BEM:time_lapse_qavg}(a,c). While the charge and shape perturbations are sinusoidal at short times in agreement with the linear theory, they depart from perfect sine waves as nonlinear effects start becoming important. Nevertheless, the underlying symmetry between the two interfaces is maintained in the non-linear regime, with $\xi(x)=-\xi_l(x+\lambda/2)$ for in-phase modes and $\xi_u(x)=-\xi_l(x)$ for anti-phase modes, where $\lambda=2\pi/k$ is the wavelength of the initial perturbation. 

 \begin{figure}
	\centering
	\adjustbox{trim={0.0\width} {0.0\height} {0.0\width} {0.0\height},clip}
	{\includegraphics[width=0.95
		\textwidth, angle=0]{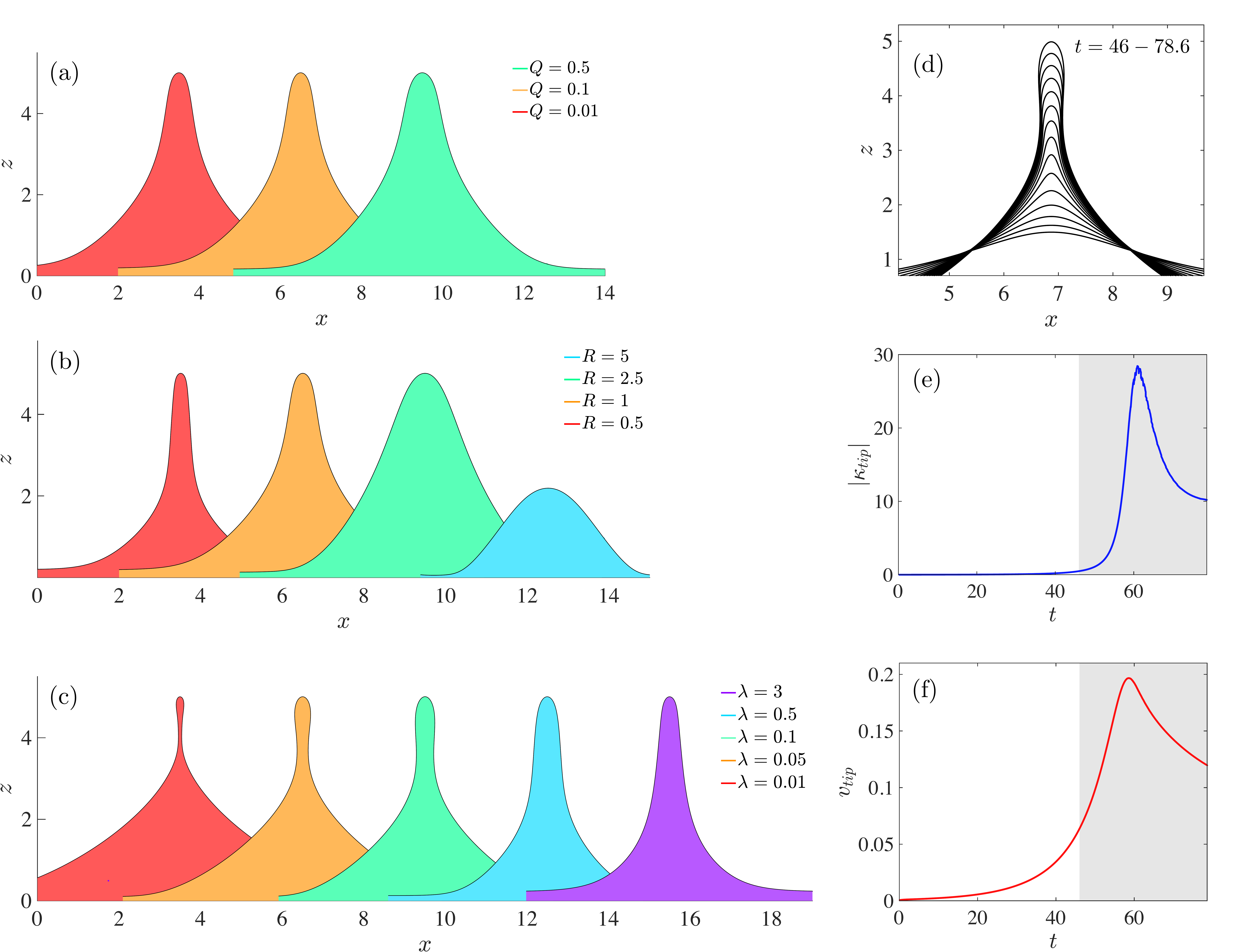} }	
	\caption{Snapshots of the tip-streaming jets (left column) formed on the upper interface during antiphase instability with different: (a) permittivity ratios in system $\mathbi{S4}$, (b) conductivity ratios in system $\mathbi{S3}$, and (c) viscosity ratios  in system $\mathbi{S5}$. Deformation parameter is $\mathcal{D}=5$ for all cases except for $R=5$ in (b), where the minimum thickness of the film reached the local grid size at $\mathcal{D}=2.18$ (disintegration from the base). The evolution of (d) the jet profile, (e) tip curvature, and (f) the vertical tip velocity is shown in the right column for system $\mathbi{S5}$ when $\lambda=0.1$. Also see videos in the Supplemental Material.}  
	\label{figs:BEM:antiphase:shape_album_&_tipstreaming_timelapse}  
\end{figure}

Further development of the flow and charge dynamics in the antiphase mode gives rise to the formation of pointed conical structures on each interface, which may eventually turn into tip streaming jets intruding the suspending phase. Initially, as new charges are brought to the tip via Ohmic conduction, the electrostatic pressure $p_E$ increases thus accelerating the fluid along the normal direction. The tip curvature increases as a result, leading to larger capillary pressures. The time scale of the induced EHD flow continues to decrease until it eventually becomes comparable and finally smaller than the charge relaxation time $\tau_c$. Consequently, the dominant mode of charge transport switches from Ohmic conduction to charge convection. Finally, the intruding jet becomes smoother on the tip and the structure begins to diverge from its conical geometry. It is evident from Figs.~\ref{figs:BEM:antiphase:shape_album_&_tipstreaming_timelapse}(e,f) that the tip curvature and velocity increase simultaneously until they peak and start to decrease following the emergence of tip formation. 

Tip streaming has also been observed in other configurations when leaky dielectric films or drops are subject to sufficiently strong electric fields \cite{collins2008tipstreaming_Nature, collins2013PNAS_scalinglaws_droptipstreaming}. Previous studies suggest that tip streaming is a local phenomenon and is nearly independent of the boundary conditions and global scales in the problem such as the thickness of the film and the strength of the electric field \cite{ganan1997current, cherney1999EHD_liquidmenisci, collins2008tipstreaming_Nature, collins2013PNAS_scalinglaws_droptipstreaming}. Accordingly, dimensional analysis for our system suggests that the jet structure and its dynamics depend only on $R$, $Q$ and $\lambda$. Figure \ref{figs:BEM:antiphase:shape_album_&_tipstreaming_timelapse}(a-c) provides a shape diagram of the tip streaming jets and illustrates how the mismatch in material properties affects the resulting morphologies. Snapshots of the jet profiles are shown at a fixed value of the deformation parameter $\mathcal{D}$, which we define as the maximum vertical deflection along the film normalized by the initial half thickness:
\begin{equation}\label{eq:BEM:deform_param}
\mathcal{D}=\max{\left|~\dfrac{z}{h}~\right| }, \qquad \text{for~all}~\bm{x}\in S.
\end{equation}
Note that our boundary element simulations are unable to capture topological changes occurring during breakup into droplets. Instead, if the thickness of the film or jet decreases below the local grid size, we denote this as a disintegration event. We predict two disintegration scenarios based on our results: pinching into droplets from the tip such as in system $\mathbi{S5}$ when $\lambda=0.01$, and breakup from the base of the jets such as in figure $\mathbi{S4}$ when $R=2.5$. A strong thinning at the base of the jet inhibits larger vertical deformations in some cases such as in $\mathbi{S4}$ when $R=5$. More information on the time evolution of the tip streaming jets shown in Fig.~\ref{figs:BEM:antiphase:shape_album_&_tipstreaming_timelapse} is included in appendix \ref{appendix:nonlinear_regime}. See Supplemental Material for videos showing the evolution of the film and the emergence of tip streaming jets in different systems with antiphase instability.

A fundamental challenge for the numerical simulations of tip streaming is the large discrepancy in the length scales present in the problem. The computational domain needs to be large enough so that the dynamics associated with small wavenumbers are accurately captured. This increases the computational cost for a given global grid resolution.  In addition, in order to capture the large local deformations during tip streaming, a high local resolution must also be  maintained throughout the simulations. Employing adaptive grid refinement allows us to address this challenge. Finally, we note that despite some similar features,  the tip streaming observed here differs from the conic cusping singularity observed in inviscid perfect conducting or perfect dielectric liquids where the tip curvature diverges in a finite time \cite{zubarev2001prfct_conductor, zubarev2002selfsimilar_dielectric}. The presence of tangential electric stresses in leaky dielectric liquids is the main difference between the two types of EHD instabilities.
%suggests that the jet profile $\xi_{jet}=z/(\mu \gamma^{-1})$ and its evolution is :
%	\begin{equation}\label{eq:BEM:dim_analysis_jet}
%	\dfrac{z^*}{\mu/\gamma}=\mathcal{F}(R,Q,\lambda,t)
%	\end{equation}
% where $z^*$ could be any spatial feature of the tip-streaming jet such as thickness of the jet or the position of its tip. 

\begin{figure}
	\centering
	\adjustbox{trim={0.0\width} {0.0\height} {0.0\width} {0.0\height},clip}
	{\includegraphics[width=0.95
		\textwidth, angle=0]{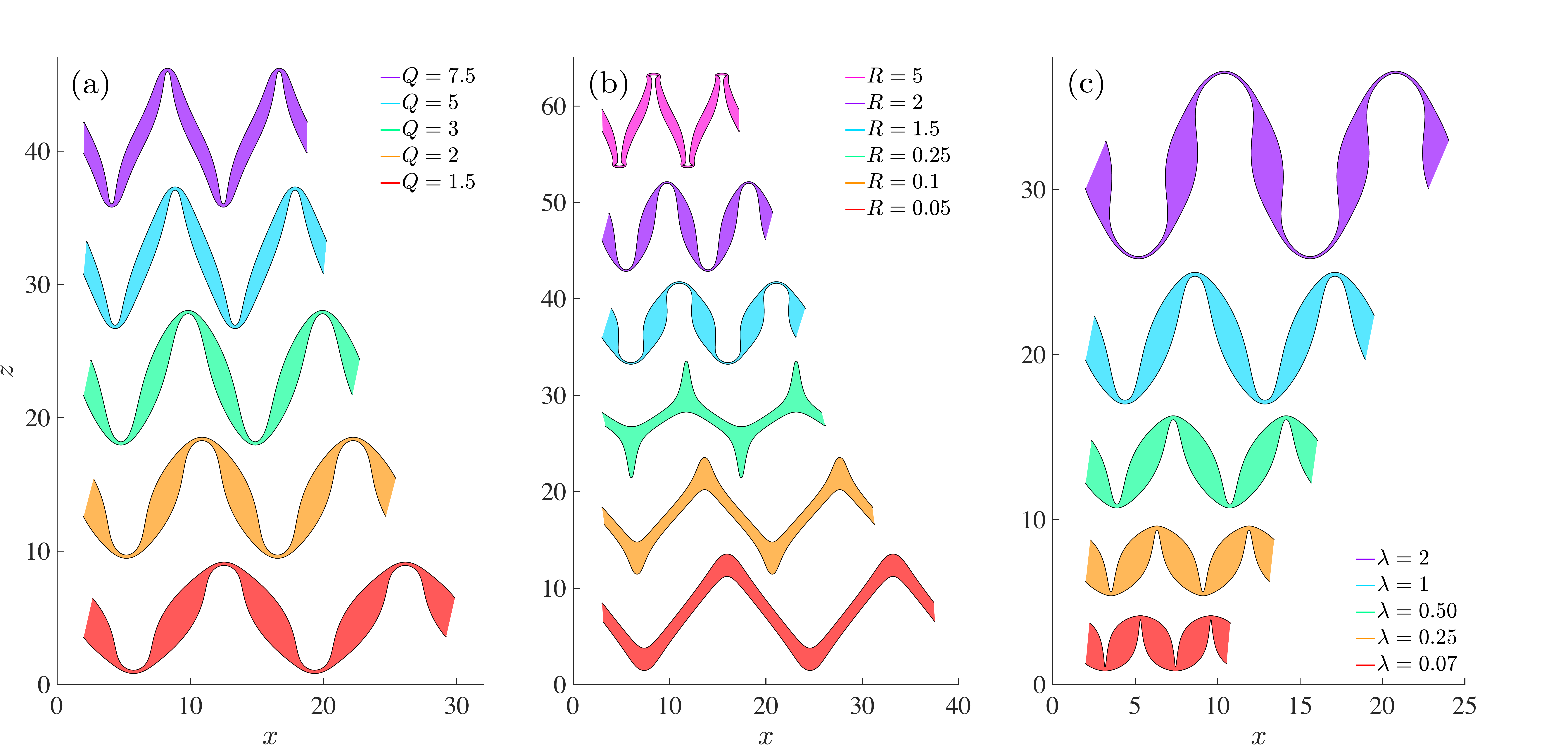} }	
	\caption{Snapshots of the films during in-phase mode instability with different: (a) permittivity ratios in system $\mathbi{S4}$, (b) conductivity ratios in system $\mathbi{S6}$, (c) viscosity ratios in system $\mathbi{S7}$. For the cases with inward jets, the minimum thickness of the film is $d=0.2$, or one tenth of the equilibrium thickness. $\mathcal{D}=5$ for outward jets. Also see videos in the Supplemental Material.  }  
	\label{figs:BEM:inphase:shape}  
\end{figure}

During the in-phase mode instability, the system undergoes a different dynamics as a result of the coupling between the flow and charge evolution. We observe two main dynamical behaviors in these systems that are shown in Fig.~\ref{figs:BEM:inphase:shape}. In systems where the equilibrium electric stresses are compressive on the film ($RQ>1$), further development of the flow and charge dynamics results in the formation of conic structures that intrude the film. These inward jets exhibit a similar evolution pattern to the ones observed in Fig.~\ref{figs:BEM:antiphase:shape_album_&_tipstreaming_timelapse}(e,f) during tip streaming. The film may breakup into droplets when these inward jets reach the other interface. On the other hand, in cases where the equilibrium electric stresses are extensional ($RQ<1$), the system develops jets that are flowing outward from the film. Generally, the dynamics in this mode is dominated by strong local effects in the nonlinear regime, which is analogous to the antiphase mode. Therefore, we can infer that the same dimensional analysis holds for the evolution of the resulting structures in this mode. Figure \ref{figs:BEM:inphase:shape} shows how the mismatch in material properties affects the behavior of the system and the resulting structures during in-phase instabilities. More information on the time evolution of these systems is included in appendix \ref{appendix:nonlinear_regime}. Videos of simulations for different systems undergoing in-phase instability with emerging inward and outward jets are included in the Supplemental Material.

\section{Concluding remarks \label{sec:conclusion}}

We have presented a theoretical and numerical model in two dimensions  to study the dynamics of a suspended viscous film that is subject to a perpendicular electric field. We performed a linear stability analysis using a charge transport model that accounts for Ohmic conduction, charge convection and finite charge relaxation, which was shown to recover the previous results of Zhang \textit{et al.}~\citep{zhang_lin2011JFM} in the limit of instantaneous charge relaxation. Two main modes of instability were identified, which are referred to as \textit{in-phase} and \textit{antiphase} modes. The system exhibits different dynamical behaviors in each mode. We characterized the effect of the relevant non-dimensional groups on the stability of the system in each mode. Our results suggest that interfacial charge convection by the flow, which had been neglected in previous related studies in the literature \cite{zhang_lin2011JFM, Zahn_Reddy2006exp_theo_channel,michael_Oneill_1970_EHD_3layer}, plays an important role in determining the dynamics of the system. Besides its destabilizing effect, it was shown that charge convection can also alter the dominant mode of instability. 

Our theoretical analysis was complemented by numerical simulations using the boundary element method so as to explore the dynamics of the system far from equilibrium. We demonstrated how the coupling of flow and interfacial charge dynamics in the antiphase mode gives rise to strongly nonlinear effects such as the formation of tip streaming jets that are intruding the suspending phase. During the in-phase mode, however, the system can undergo various nonlinear routes depending on the type of electric stresses in the base state. In cases where the equilibrium electric stresses are compressive on film, we observed the emergence of inwards jets that are drawn towards the film on each interface. Conversely, if the equilibrium electric stresses are extensional, the system was shown to develop conical jets that are flowing outward from the film. Strong local effects are a key common feature between the two modes during their non-linear regime of growth. Finally, we characterized the effect of different controlling parameters on the dynamical behavior of the system and the resulting structures in the non-linear regime.

The present study casts new light on the dynamics of suspended viscous films under applied electric fields, where the interplay between electric, hydrodynamic and capillary forces can result in a plethora of dynamical behaviors. Our knowledge on the fastest growing modes along with the additional insight into the nonlinear behavior of the system should be of great use for engineering applications where EHD instabilities are exploited to create patterns or structures with prescribed morphologies and length scales. Since the mentioned nonlinear phenomena such as tip streaming are inherently three dimensional, extending this work to describe three-dimensional pattern formation would be of interest, as would including the effects of fluid inertia, which may play a role during regimes of rapid nonlinear growth.

\begin{acknowledgments}
	The authors thank Petia Vlahovska and Michael Miksis for useful conversations, and gratefully acknowledge funding from National Science Foundation Grant CBET-1705377.
\end{acknowledgments}

\appendix

\section{Linear stability analysis}\label{appendix:lin_stab}
In the base state, all liquid layers are at rest, both interfaces have flat shapes, $\tilde{\xi}_{u}=-\tilde{\xi}_{l}=1$, and the electric potential field in each layer reads:
\begin{align}
&\tilde{\varphi}_u(z)= -z +(1-R),&  z\ge1, \\
&\tilde{\varphi}_m(z)= -Rz, &  -1\le z \le 1, \\
&\tilde{\varphi}_l(z)= -z-(1-R),&  z\le -1, 
\end{align}
for which
\begin{eqnarray}
   & \tilde{\bm{E}}_l= \tilde{\bm{E}}_u=R^{-1}\tilde{\bm{E}}_m=\left(0, 1 \right), &\\
   & \tilde{q}_{u}=-\tilde{q}_{l}=(1-RQ), &\\
&    \tilde{p}_u-\tilde{p}_m=\tilde{p}_l-\tilde{p}_m=(1-QR^2)/2.&
\end{eqnarray}

After perturbing the base state, we linearize the governing equations and boundary conditions. Next, we seek normal-mode solutions of the form  $\varphi'(x,z,t)=\hat{\varphi}(z)\exp{(st+ikx)}$, with similar expressions for all the variables. Substituting the normal-modes into the governing equations and using the decay properties as $z\rightarrow \pm \infty$ provides the amplitude of the normal modes as:
\begin{align}
&\hat{\varphi}_u (z) = A_u e^{-kz}, & \qquad  z\ge1, \\
&\hat{\varphi}_m (z) = A_{m1} e^{-kz}+A_{m2} e^{kz}, &  \qquad -1\le z \le 1, \\
&\hat{\varphi}_l (z) = A_l e^{kz}, & \qquad z\le -1, 
\end{align}
\begin{align}
&\hat{p}_u(z) =   B_u e^{-kz},& \qquad z\ge1, \\ 
&\hat{p}_m(z) = B_{m1} e^{-kz}+B_{m2} e^{kz},  & \qquad -1\le z \le 1, \\
&\hat{p}_l(z) = B_l e^{kz}, & \qquad z\le -1, 
\end{align}
\begin{align}
& \hat{u}_u(z)=  C_u e^{-kz} -i B_u ( \dfrac{1+\lambda}{2} ) z e^{-kz}, &  z\ge1, \\
& \hat{u}_m(z)= C_{m1} e^{-kz}+C_{m2} e^{kz}-i B_{m1}(\dfrac{ 1+\lambda^{-1}}{2})z e^{-kz}+i B_{m2} (\dfrac{ 1+\lambda^{-1}}{2} ) z e^{kz},  &  -1\le z \le 1, \\
& \hat{u}_l(z)=C_l e^{kz}+ i B_l(\dfrac{1+\lambda}{ 2})z e^{kz},& z\le -1,
\end{align}
\begin{align}
& \hat{w}_u(z)= D_u e^{-kz}+ B_u (\dfrac{1+\lambda}{2})ze^{-kz},  & z\ge1, \\
& \hat{w}_m(z)= D_{m1} e^{-kz}+D_{m2} e^{kz}+ B_{m1}(\dfrac{1+\lambda^{-1}}{2})z e^{-kz}+ B_{m2}(\dfrac{1+\lambda^{-1}}{2})z e^{kz},  & -1\le z \le 1, \\
& \hat{w}_l(z)= D_l e^{kz}+ B_l (\dfrac{1+\lambda}{2} ) ze^{kz}, & z\le -1, 
\end{align}
where
\begin{align}
&D_u = i C_u +B_u (1+\lambda)(2k)^{-1}, && D_l = -i C_l -B_l (1+\lambda)(2k)^{-1},\\
&D_{m1} =  iC_{m1} +B_{m1} (1+\lambda^{-1})(2k)^{-1}, &&  D_{m2} = -i C_{m2} -B_{m2}(1+\lambda^{-1})(2k)^{-1}.
\end{align}
Applying the boundary conditions for the perturbation variables results in a linear algebraic system for the unknown coefficients $A_j$, $B_j$, $C_j$ and $D_j$ where $j\in \{l,m1,m2,u\}$. Finally, we obtain the dispersion relation by setting the determinant of the algebraic system to zero.

\section{Fastest growing mode}\label{appendix:kmax_plots}
\begin{figure}[h!]	
	\centering
	\adjustbox{trim={0.0\width} {0.0\height} {0.0\width} {0.0\height},clip}
	{\includegraphics[width=.80
		\textwidth, angle=0]{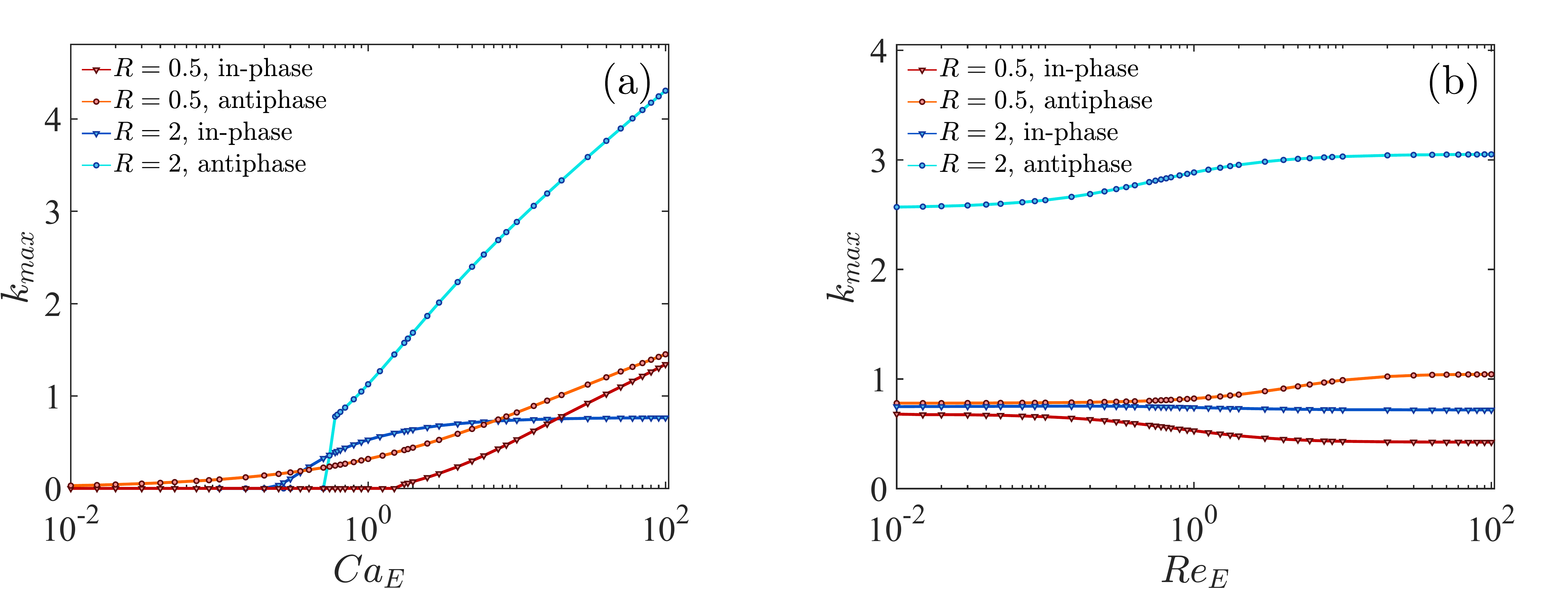}}
	\caption{Fastest-growing wavenumber in each mode of instability as a function of: (a) electric capillary number, and (b) electric Reynolds number for two different systems with $R=0.5$ and $R=2$ with $(Q,\lambda)=(1,1)$. $Re_E=1$ in (a) and $Ca_E=10$ in (b).}  
	\label{figs:appendix:kmax_Ca_Re}  
\end{figure}

\begin{figure}[ht!]	
	\centering
	\adjustbox{trim={0.0\width} {0.0\height} {0.0\width} {0.0\height},clip}
	{\includegraphics[width=1
		\textwidth, angle=0]{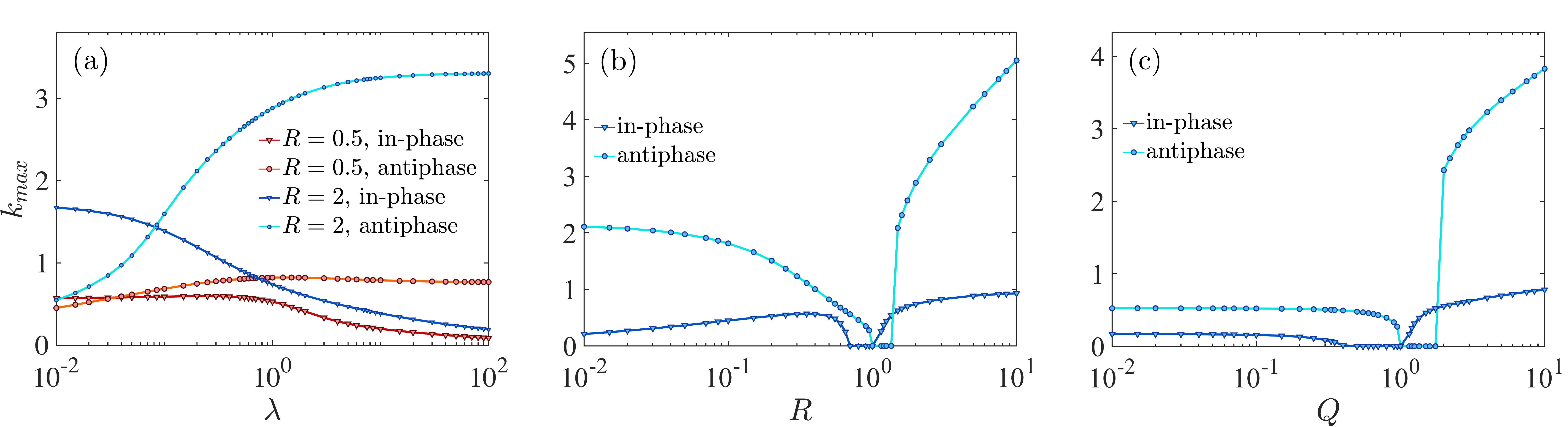} }
	\caption{ Fastest-growing wavenumber in each mode of instability as a function of:  (a) viscosity ratio $\lambda$ for two different systems with $R=0.5$ and $R=2$, (b) conductivity ratio $R$, and (c) permittivity ratio $Q$. The permittivity ratio is set to $Q=1$ in (a,b), the conductivity ratio is $R=1$ in (c) and $\lambda=1$ in (b,c). In all systems $(Ca_E,Re_E)=(10,1)$.  }  
	\label{figs:appendix:kmax_R_Q_lambda}  
\end{figure}

The maximum growth rate and the corresponding wavenumber in each mode are defined respectively as $s_{max}$ and $k_{max}$. Figures \ref{figs:appendix:kmax_Ca_Re} and \ref{figs:appendix:kmax_R_Q_lambda} show how the non-dimensional parameters governing the system affect $k_{max}$ in each mode. 
%\begin{figure}[ht!]	
%	\centering
%	\adjustbox{trim={0.0\width} {0.0\height} {0.0\width} {0.0\height},clip}
%	{\includegraphics[width=0.35
%		\textwidth, angle=0]{./linear_stability/NEW_KBC/kmax_lambda_7March2021} }
%	\caption{Fastest-growing wavenumber in each mode of instability as a function of viscosity ratio $\lambda$ for two different systems with $R=0.5$ and $R=2$ while $(Q,\lambda,Ca_E)=(1,1,10,1)$.}  
%	\label{figs:appendix:kmax_lambda}  
%\end{figure}

\clearpage

\section{Nonlinear regime}\label{appendix:nonlinear_regime}
\begin{figure}[h!]
	\centering
	\adjustbox{trim={0.0\width} {0.0\height} {0.0\width} {0.0\height},clip}
	{\includegraphics[width=0.85
		\textwidth, angle=0]{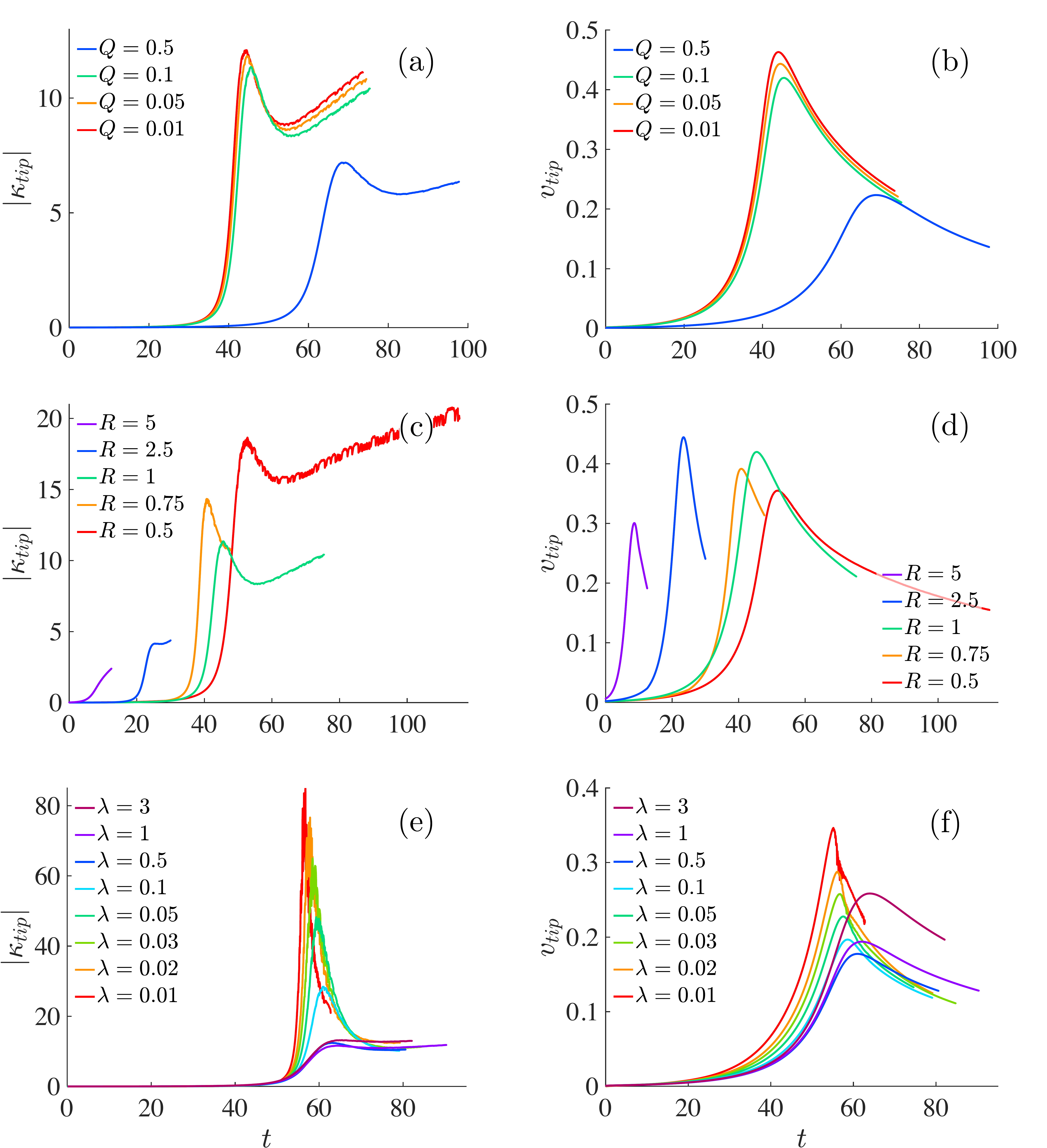} }	
	\caption{Evolution of the tip streaming jets during antiphase instability as a function of: (a,b) permittivity ratio in system $\mathbi{S4}$, (c,d) conductivity ratio in system $\mathbi{S3}$, and (e,f) viscosity ratio  in system $\mathbi{S5}$. Left and right columns show the evolution of the tip curvature and vertical tip velocity, respectively. }  
	\label{figs:BEM_appendix:anti:curvature_vtip}  
\end{figure}

\begin{figure}[h!]
	\centering
	\adjustbox{trim={0.0\width} {0.0\height} {0.0\width} {0.0\height},clip}
	{\includegraphics[width=0.85
		\textwidth, angle=0]{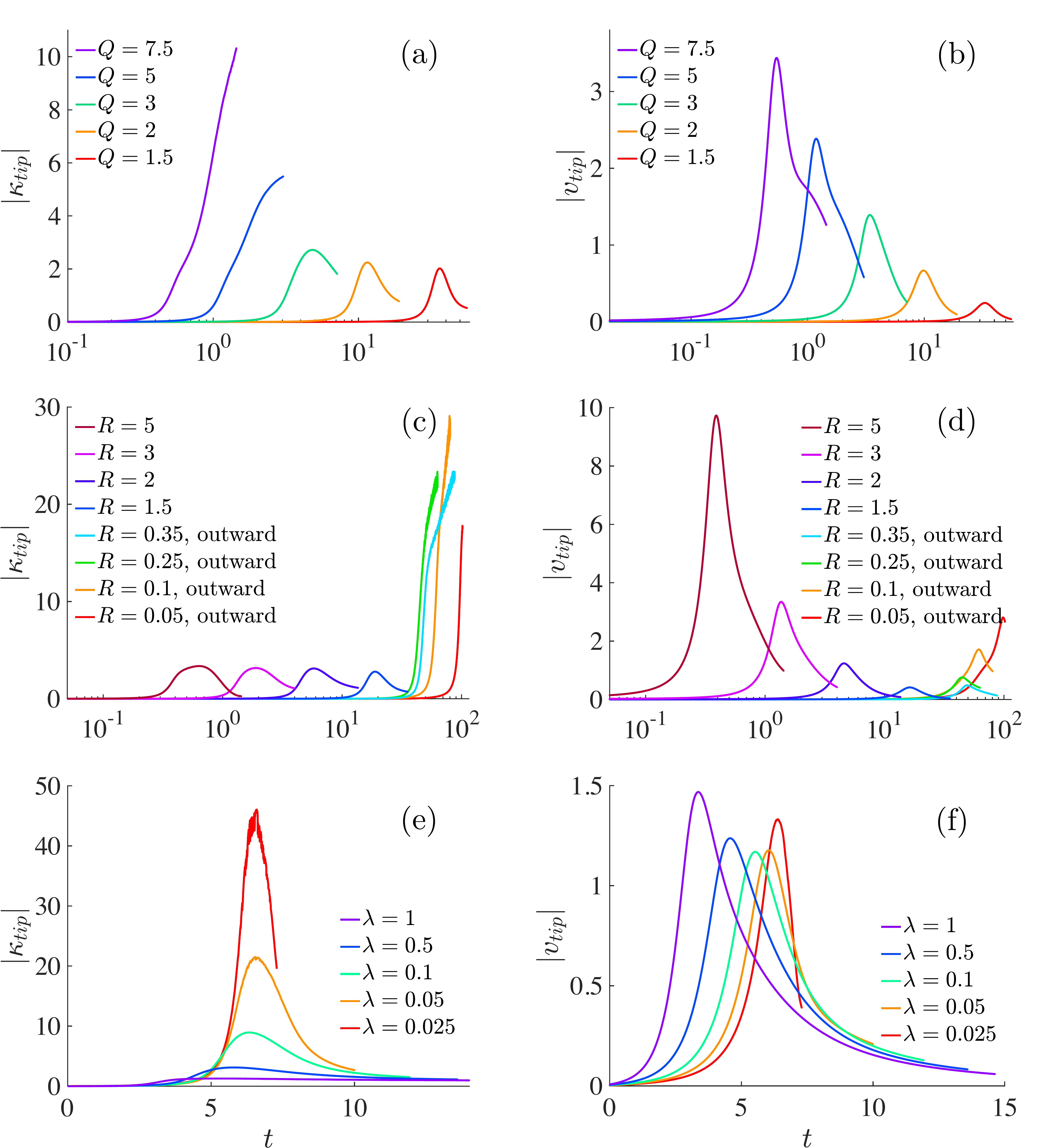} }	
	\caption{Evolution of the inward and outward jets during in-phase instability as a function of: (a,b) permittivity ratio in system $\mathbi{S4}$, (c,d) conductivity ratio in system $\mathbi{S6}$, and (e,f) viscosity ratio  in system $\mathbi{S7}$. Left and right columns show the evolution of the tip curvature and vertical tip velocity, respectively. }  \vspace{-0.4cm}
	\label{figs:BEM_appendix:inphase:curvature_vtip}  
\end{figure}

The development of the flow and charge dynamics in antiphase instabilities gives rise to the formation of tip streaming jets during the nonlinear regime. On the other hand, during in-phase instability, we observe inward and outward jets that intrude the film and the suspending phase, respectively. Figures \ref{figs:BEM_appendix:anti:curvature_vtip} and \ref{figs:BEM_appendix:inphase:curvature_vtip} show the effect of the mismatch in the different material properties on the evolution of the emerging jets in each mode.

%\subsection{Antiphase mode}\label{appendix:nonlinear_regime:antiphase_mode}

\clearpage

%\subsection{In-phase mode}\label{appendix:nonlinear_regime:inphase_mode}

\bibliography{Refs_PRF_shortjrtitle}

\end{document}